\documentclass[aip,reprint,citeautoscript]{revtex4-1}
\usepackage{amssymb}
\usepackage{amsmath}
\usepackage{graphicx}
\usepackage{color}
\usepackage{todonotes}
\usepackage{braket}
\usepackage{algorithm}
\usepackage{algpseudocode}

\newcommand{\Dt}{\Delta t}

\newcommand{\trajarg}{\left[\bf{X}_{t,t'}\right]}
\newcommand{\trajargT}{\left[\bf{X}_{0,T}\right]}

\newcommand{\gradt}{\nabla_\theta}
\newcommand{\gradp}{\nabla_\psi}
\newcommand{\DKL}{D_{\mathrm{KL}}(p_\theta|p_{s})}
\newcommand{\dKL}{d_{\mathrm{KL}}(p_\theta|p_{s})}
\renewcommand{\eqref}[1]{Eq.~\ref{#1}}
\renewcommand{\bf}[1]{\mathbf{#1}}
\newcommand{\G}{{\mathbb G}}

\definecolor{blue2}{rgb}{0.0, 0.3, 0.7}
\usepackage[colorlinks = true, citecolor = blue2, linkcolor = blue2, urlcolor=blue2]{hyperref}

\begin{document}

\title{Reinforcement learning of rare diffusive dynamics}
\author{Avishek Das}\thanks{These authors contributed equally}\email{avishek\_das@berkeley.edu}
\affiliation{Department of Chemistry, University of California, Berkeley CA 94609 \looseness=-1}
\author{Dominic C. Rose}\thanks{These authors contributed equally}\email{dominic.rose1@nottingham.ac.uk}
\affiliation{School of Physics and Astronomy, University of Nottingham,  Nottingham NG7 2RD, United Kingdom \looseness=-1}
\affiliation{Centre for the Mathematics and Theoretical Physics of Quantum Non-Equilibrium Systems,  University of Nottingham, Nottingham NG7 2RD, United Kingdom \looseness=-1}
\author{Juan P. Garrahan}
\email{juan.garrahan@nottingham.ac.uk}
\affiliation{School of Physics and Astronomy, University of Nottingham,  Nottingham NG7 2RD, United Kingdom \looseness=-1}
\affiliation{Centre for the Mathematics and Theoretical Physics of Quantum Non-Equilibrium Systems, University of Nottingham, Nottingham NG7 2RD, United Kingdom \looseness=-1}
\author{David T. Limmer} 
\email{dlimmer@berkeley.edu}
\affiliation{Department of Chemistry, University of California, Berkeley CA 94609 \looseness=-1}
\affiliation{Kavli Energy NanoScience Institute, Berkeley, CA 94609 \looseness=-1}
\affiliation{Materials Science Division, Lawrence Berkeley National Laboratory, Berkeley, CA 94609 \looseness=-1}
\affiliation{Chemical Science Division, Lawrence Berkeley National Laboratory, Berkeley, CA 94609\looseness=-1}

\begin{abstract}
We present a method to probe rare molecular dynamics trajectories directly using reinforcement learning. We consider trajectories that are conditioned to transition between regions of configuration space in finite time, like those relevant in the study of reactive events, as well as trajectories exhibiting rare fluctuations of time-integrated quantities in the long time limit, like those relevant in the calculation of large deviation functions. In both cases, reinforcement learning techniques are used to optimize an added force that minimizes the Kullback-Leibler divergence between the conditioned trajectory ensemble and a driven one. Under the optimized added force, the system evolves the rare fluctuation as a typical one, affording a variational estimate of its likelihood in the original trajectory ensemble. Low variance gradients employing value functions are proposed to increase the convergence of the optimal force. The method we develop employing these gradients leads to efficient and accurate estimates of both the optimal force and the likelihood of the rare event for a variety of model systems.
\end{abstract}

\maketitle

\section{Introduction}
Rare but important events play a significant role in phenomena occurring throughout the sciences, ranging from physics \cite{Touchette2009} and chemistry \cite{chandler1998barrier}, to climate science \cite{Webber2019} and economics.\cite{stanley2007economic} As a consequence methods developed to study rare events can transcend disciplines. In molecular systems, rare events determine the rates by which chemical reactions occur and  phases interconvert,\cite{peters2017reaction} and they also encode the response of systems driven to flow or unfold.\cite{gao2017transport,gao2019nonlinear,limmer2021large,kuznets2021dissipation,noe2009constructing} Strategies that afford a means of studying rare dynamical events in statistically unbiased ways are particularly desired, in order to deduce the intrinsic pathways by which they occur and to evaluate their likelihoods. Borrowing notions from reinforcement learning,\cite{Sutton2018} we have developed a method to generate rare dynamical trajectories directly through the optimization of an auxiliary dynamics that generates an ensemble of trajectories with the correct relative statistical weights. Within this ensemble of trajectories, a variational estimate of the likelihood of the rare event is obtainable from a simple expectation value. 

Much research has been devoted to the enhanced sampling of molecular dynamics simulations, yet there remains active areas of open research. Methods for sampling dynamical fluctuations, especially those away from equilibrium, are considerably less developed then their equilibrium and configurational counterparts.\cite{frenkel2001understanding,Zhang2019} Recent work has sought to construct methods for finding an effective auxiliary dynamics,\cite{Nemoto2016,Oakes2020,Whitelam2020,Kappen2016,Ray2018,Ferre2018,Zhang2021} with the goal of sampling rare dynamical fluctuations with the corresponding correct statistical weights directly, by evolving simulations with additional parametrized forces. Such methods are often designed to approximate the so-called Doob transform\cite{Chetrite2014,Jack2015a,Chetrite2015,Causer2021} which is the unique force that evolves a trajectory conditioned on a rare event. 

A general approach to the optimization of a sampling dynamics based on a variational principle for the Doob transform for diffusive processes has recently been developed \cite{Das2019}. Within this context of diffusive processes, optimal forces have been used to elucidate mechanisms and rates of nonlinear response,\cite{grandpre2018current,tociu2019dissipation} to encode dynamical phase diagrams,\cite{GrandPre2020,Nemoto2019,Keta2020} and to deduce inverse design principles.\cite{Das2021,pineros2021inverse} In this work we aim to extend a reinforcement learning\cite{Sutton2018} based approach to the optimization of a sampling dynamics to diffusive systems, building on the work of Refs. \onlinecite{Das2019,Rose2020} and past literature on reinforcement learning for continuous time processes \cite{Munos1998,Munos2005,Doya2000,Bradtke1994,Vamvoudakis2010,Fremaux2013,Beard1997,AbuKhalaf2005}.

The techniques of reinforcement learning aim to learn the best decisions to make in each state in order to achieve some goal. Algorithms developed in this context have led to many significant advancements in recent years across tasks requiring an intelligent agent to interact with an environment, such as in gameplay\cite{Mnih2015,Vinyals2019,Silver2018} and robotics \cite{Haarnoja2018,Haarnoja2018a,OpenAI2019}, with a variety of recent applications in physics \cite{Bukov2018,Bukov2018a,Yao2020,Fosel2018,Chen2019,Dalgaard2020,Barr2020,Gispen2020}.
However, many of these situations are framed as discrete time problems, with relatively little work done in stochastic continuous time control \cite{Munos1998,Munos2005}.
For diffusive processes and importance sampling molecular dynamics, we formulate a reinforcement learning procedure to learn the correct force to influence the probability of choosing each next state. From this perspective, we take a policy gradient based approach \cite{Sutton2000,Marbach2003,Munos2005,Haarnoja2018,Haarnoja2018a}, learning a generative model for the evolution of the state.
The optimized force found is such that rare events are made typical while staying close to the original force, providing a dynamics that can aid in efficiently sampling the targeted trajectory ensemble.

A key advantage of the reinforcement learning techniques we develop is the use of an additional learning process for a function which guides the optimization of the dynamics, a so-called value function \cite{Sutton1988}, which describes how relevant each state is to the rare events of interest.
This value function substantially reduces the variance in estimates of the gradient of the parameters specifying a force, allowing for the use of less data in each optimization step and subsequently more complex approximations to the auxiliary dynamics.
We show how this approach can be successfully applied to both finite time problems in which the dynamics is constrained to guarantee the occurrence of some rare transition like a barrier crossing, and to time-homogeneous problems where we are interested in the statistics of time-integrated observables in the long time limit as characterized by its large deviation function.

\section{Trajectory ensemble Formalism}
We consider systems evolving with a diffusive dynamics over time $t$ of a configuration $\bf{x}$.
These configurations evolve according to a force vector $\bf{F}(\bf{x},t)$ and noise vector of equal dimension $\bf{W}$ with associated constant noise matrix $\G$ invertible within the stochastically evolving subspace, represented by the stochastic differential equation (SDE)
\begin{eqnarray}\label{eq:sde}
	d\bf{x}=\bf{F}(\bf{x},t)dt+\G \cdot d\bf{W},
\end{eqnarray}
where the noise $\bf{W}$ follows a Wiener process, with increments $d\bf{W}$ drawn from a Gaussian with zero mean and $dt$ variance.Throughout we will work in dimensionless variables that imply unit energy scales and mobilities.
The requirement of $\G$ being invertible within the stochastic subspace may in principle be relaxed, however in that case there may be multiple noise vectors corresponding to the same change of state, making the evaluation of transition probabilities  necessary for our optimization approach difficult. We will follow the Ito convention for ease of notation and implementation with standard numerical integrators. Throughout, we do not assume in Eq.~\ref{eq:sde} that the force is gradient or that the noise obeys a detailed balance, and thus our approach is generally applicable to equilibrium as well as nonequilibrium dynamics. 
 
We aim to probe rare fluctuations in trajectory observables. Here we consider trajectories, $\bf{X}_{0,T}$, defined as the sequence of configurations over an observation time $T$, though generalizations of fluctuating observation times are possible.\cite{budini2014fluctuating}  Generally, we will consider observables that are functions of time-integrated variables over the trajectory,
\begin{eqnarray}\label{eq:time-integrated-observable}
	O\trajargT=\int_0^Tdt \, A[\bf{x}_{t},t]+ \bf{B}[\bf{x}_{t},t]\cdot \dot{\bf{x}}(t),
\end{eqnarray}
where the first term is a state dependent observable, while the second term depends on a stochastic increment, with both $A[\bf{x}_{t},t]$ and $\bf{B}[{\bf{x}}_{t},t]$ being state dependent. However, we will also consider cases in which $A[\bf{x}_{t},t]$ is a function of a single time in order to impose end point conditioning.
Expectations of functions of such observables are defined through path integrals of the form
\begin{eqnarray}
	\left\langle f\left(O\left[\bf{X}_{t,t'}\right]\right)\right\rangle_p
	=\int D \bf{X}_{t,t'}d \bf{x}_t \, P\left[\bf{X}_{t,t'}\right]f\left(O\left[\bf{X}_{t,t'}\right]\right),
\end{eqnarray}
where $P\left[\bf{X}_{t,t'}\right]$ is the total probability of a trajectory decomposable into $P\left[\bf{X}_{t,t'}\right]=p\left[\bf{X}_{t,t'} | \bf{x}_t\right] \rho(\bf{x}_t)$ where $p\left[\bf{X}_{t,t'} | \bf{x}_t\right]$ is the transition probability conditioned on starting in configuration $\bf{x}_t$ with initial probability $\rho(\bf{x}_t)$.

Probabilities for trajectories between times $t$ and $t'$ starting at $\bf{x}_t$  are defined by
\begin{eqnarray}\label{eq:trajectory-prob}
	p\left[\bf{X}_{t,t'}|\bf{x}_t\right]
	&\propto\exp\left \{-\frac{1}{2}\int_t^{t'}dt''\left |\G^{-1} \cdot \left(\dot{\bf{x}} - \bf{F}\right)\right |^2\right\}
\end{eqnarray}
where we suppressed the arguments of $\bf{x}_{t}$ and $\bf{F}[\bf{x}_{t},t]$ for shorthand. This is the standard Onsager-Machlop form for the diffusive dynamics considered here.\cite{taniguchi2007onsager} The measure over paths between times $t$ and $t'$ starting from position $\bf{x}_t$ is defined such that 
\begin{eqnarray}
	\int D\bf{X}_{t,t'}p\left[\bf{X}_{t,t'}| \bf{x}_t\right]=1
\end{eqnarray}
where the transition probability is normalized when integrated over all trajectories.
These path probabilities satisfy
\begin{eqnarray}
	p\left[\bf{X}_{t,t''}|\bf{x}_{t}\right]=p\left[\bf{X}_{t',t''}|\bf{x}_{t'}\right]p\left[\bf{X}_{t,t'}|\bf{x}_{t}\right]
\end{eqnarray}
and 
\begin{equation}
	D\bf{X}_{t,t''}=D\bf{X}_{t',t''}D\bf{X}_{t,t'}
\end{equation}
due to the Markovian noise in Eq.~\ref{eq:sde}.

Trajectories sampled with  $P\left[\bf{X}_{0,T}\right]$ will be dominated by the most typical values of $O\trajargT$. We will encode the rare trajectories with atypical values of $O\trajargT$  by reweighting the original trajectory ensemble defined by \eqref{eq:trajectory-prob}, multiplying each trajectory by an observable dependent factor. Such reweightings occur naturally in statistical studies of rare events and are isomorphic to extended ensemble approaches in equilibrium configurational problems. 
The ensemble of events we are interested in is constructed by weighting the probability of trajectories in the original dynamics by an exponentially positive number,
\begin{eqnarray}
\label{eq:reweighted-trajectory-prob}
		P_s\trajargT=e^{-sO\trajargT-\lambda(s,T)}P\trajargT,
	\end{eqnarray}
 where $P_s\trajargT$ is denoted as a tilted path ensemble, biased by a statistical field $s$ in such a way to promote rare fluctuations in $O\trajargT$. The quantity $\lambda(s,T)$ normalizes the tilted distribution, and is identifiable as a cumulant generating function (CGF)
	\begin{eqnarray}
		\lambda(s,T)=\ln Z(s,T)=\ln\left< e^{-sO\trajargT}\right>_p \, ,
	\end{eqnarray}
and equal to the logarithm of the tilted path partition function $Z(s,T)$. The reweighted path ensemble generally defines a new transition probability $p_s\left[\bf{X}_{t,t'}|\bf{x}_t\right]$ and initial condition. The evaluation of $\lambda(s,T)$ is a common objective in studies of diffusive systems as it describes the statistics of $O\trajargT$.
	Contributions to $\lambda(s,T)$ or $P_s\trajargT$ are dominated by trajectories with large or small values of $O\trajargT$, depending on the sign of $s$. The exponential bias, $\exp(-sO\trajargT)$, can also be constructed to function as a filter based on fulfilling specific criteria.  In such cases $P_s\trajargT$ is identified as the probability that a trajectory fulfills a specific conditioning, and its ensemble a corresponding conditioned path ensemble.  Common examples are Brownian bridges \cite{Majumdar2015,grela2021non,de2021generating}, where trajectories are conditioned to end at $\bf{x}_{T}=\bf{x}'$, in which $O\trajargT$ is 1 if $\bf{x}_{T}=\bf{x}'$ and is $0$ otherwise, and $s$ is taken sufficiently negative that only those trajectories for which the constraint is satisfied have significant weight.

\section{Gradient optimization for finite time constrained dynamics}
Our aim is to find a dynamics which generates trajectories with probability as close to the reweighted trajectories ensemble as possible.
For the diffusive dynamics considered here, this is exactly achievable in principle through a so-called generalized Doob transformation.\cite{Borkar2003,Popkov2010,Jack2010,Chetrite2014,Jack2015a,Carollo2018} The generalized Doob transformation defines a modified dynamics with an added drift force that is generally time dependent but with an identical noise as in the original SDE. However, constructing this transformation is often not possible in practice, as it requires diagonalizing a modified Fokker-Planck operator which in interacting systems is exponentially complex.\cite{Causer2021} 
Here we aim to parametrize a drift force with tunable parameters $\theta$ to approximate the generalized Doob transform. With the modified force, $\bf{F}_\theta(\bf{x},t)$, we have a modified SDE
\begin{eqnarray}\label{eq:modifiedsde}
	d\bf{x}=\bf{F}_\theta(\bf{x},t)dt+\G \, d\bf{W},
\end{eqnarray}
with corresponding trajectory probabilities
\begin{eqnarray}\label{eq:parametrized-trajectory-prob}
	p_\theta\left[\bf{X}_{t,t'}|\bf{x}_t\right]
	&\propto\exp\left \{-\frac{1}{2}\int_t^{t'}dt''\left |\G^{-1} \cdot  \left(\dot{\bf{x}} - \bf{F}_\theta\right)\right |^2\right\} \, 
\end{eqnarray}
which still satisfy the Markovian properties of the original dynamics and the same normalization constant. 
See Ref.~\onlinecite{Rose2020} for a discussion of problems in which the optimal dynamics is required to be non-Markovian, in the context of discrete time Markov processes.

We seek to learn a set of parameters $\theta$ to minimize the Kullback-Leibler (KL) divergence between the modified dynamics and the reweighted trajectory ensemble defined by Eq.~\ref{eq:reweighted-trajectory-prob}. The KL divergence is defined as
\begin{eqnarray}
	\DKL
	&=\left\langle\ln\left(\frac{p_\theta\left[\bf{X}_{0,T}|\bf{x}_0\right] \rho(\bf{x}_0)}{p_s\left[\bf{X}_{0,T}|\bf{x}_0\right]\rho(\bf{x}_0)}\right)\right\rangle_{p_\theta} \, ,
\end{eqnarray}
where the expectation is taken with respect to the parametrized dynamics. This quantity is a measure of the similarity between the modified and reweighted trajectory ensembles. Achieving a zero value when $p_\theta$ is given by the generalized Doob transform, the KL divergence has a unique minimum when this Doob transformed dynamics is contained within the space of parametrized dynamics, providing a variational estimate of the CGF. We note that this definition of the KL divergence differs from much of the literature considering optimization of a parametrized diffusive dynamics,\cite{Kappen2012,Chernyak2013,Thijssen2015,Kappen2016} where the parametrized dynamics $p_\theta$ and target dynamics $p_s$ appear in an opposite way.
In principle the initial distribution should also be parametrized, as it will be modified by the reweighting, however depending on the space of distributions chosen these can be hard to sample.
We drop this modification for simplicity. 

\subsection{Low variance gradient estimation}
In order to optimize the force, $\bf{F}_\theta$, we follow techniques introduced in the reinforcement learning literature\cite{Sutton2018,Todorov2009,Neu2017,Levine2018,Haarnoja2018,Geist2019}. Substituting the parametrized and reweighted trajectory probabilities into the KL divergence, we may rewrite it as an average over a parameter dependent time-integrated observable
\begin{eqnarray}\label{eq:bound}
	\DKL
	&=-\left\langle R \trajargT \right\rangle_{p_\theta}+\lambda(s,T)
\end{eqnarray}
where in the language of reinforcement learning we define a return, $R\trajargT$, as
\begin{eqnarray}\label{eq:returnfull}
	R\left[\bf{X}_{0,T}\right]
	&=&-sO\left[\bf{X}_{0,T}\right]-\ln\left(\frac{p_\theta\left[\bf{X}_{0,T}|\bf{x}_0\right]}{p\left[\bf{X}_{0,T}|\bf{x}_0\right]}\right)\nonumber\\
%	&=&-sO\left[x_{t,t'}\right]+\Biggl\{ \frac{1}{2}\int_t^{t'}dt\left[G(t)^{-1}\left(\dot{x}(t) - F_\theta(t)\right)\right]^2\nonumber\\
%	&&-\frac{1}{2}\int_t^{t'}dt\left[G(t)^{-1}\left(\dot{x}(t) - F(t)\right)\right]^2\nonumber\Biggr\}\\
	%&=&\int_0^{T}dt \, r(\dot{\bf{x}},\bf{x},t),
\end{eqnarray}
with the negative of the average of the second term measuring the KL divergence, $D_\mathrm{KL}(p_{\theta}|p)$, between the parametrized dynamics and the original dynamics.
This return is analogous to a regularized form of reinforcement learning \cite{Neu2017,Geist2019} similar to that considered in maximum-entropy reinforcement learning \cite{Haarnoja2018,Haarnoja2018a,Levine2018}.
When evaluated at the generalized Doob transform the KL divergence vanishes and the return evaluates to the CGF. Away from the Doob transform, the positivity of the KL divergence results in the return variationally bounding the CGF from below.\cite{Chetrite2015}
%The so-called reward $r(\dot{\bf{x}},\bf{x},t)$ is the integrand found by combining the second term with the observable
%\begin{eqnarray}
%r(\dot{\bf{x}},\bf{x},t) &=& -sA[\bf{x},t]- s\bf{B}[\bf{x},t]\cdot \dot{\bf{x}}  \\
%&&+ \frac{1}{2} \left \{ | \G^{-1}\cdot\left(\dot{\bf{x}} - \bf{F}_\theta\right)| ^2 -| \G^{-1}\cdot\left(\dot{\bf{x}} - \bf{F}\right)|^2 \right \} \nonumber 
%\end{eqnarray}
%which is a local measure of the return.

We aim to minimize the KL divergence through stochastic gradient descent in the parameter space.  For this we need the gradient of $\DKL$ with respect to $\theta$,
\begin{eqnarray}\label{eq:unsupervised-gradient}
	\gradt\DKL=
	&-&\left\langle R\trajargT\gradt\ln p_\theta\left[\bf{X}_{0,T}|\bf{x}_0\right]\right\rangle_{p_\theta}\nonumber\\
\end{eqnarray}
where we note
\begin{equation}
\left\langle\gradt R\left[\bf{X}_{0,T}\right]\right\rangle_{p_\theta} = 0
\end{equation}
due to conservation of probability.\cite{Rose2020} 
%\begin{eqnarray}\label{eq:return-gradient}
%	\gradt R\left[x_{0,T}\right]=-\gradt \ln p_\theta\left[x_{0,T}|x_0\right],
%\end{eqnarray}
%and thus this term is zero, since
%\begin{eqnarray}\label{eq:path-total-deriv}
%	\left\langle\gradt \ln p_\theta\left[x_{0,T}|x_0\right]\right\rangle_{p_\theta}
%	&=&\left\langle\frac{\gradt p_\theta\left[x_{0,T}|x_0\right]}{p_\theta\left[x_{0,T}|x_0\right]}\right\rangle_{p_\theta}\nonumber\\
%	&=&\gradt\left\langle1\right\rangle_{p_\theta}=0.
%\end{eqnarray}
%We therefore have
%\begin{eqnarray}
%	\gradt\DKL=-\left\langle R\left[x_{0,T}\right]\gradt\ln p_\theta\left[x_{0,T}|x_0\right]\right\rangle_{p_\theta}.
%\end{eqnarray}
%Before further analysing this gradient we take a moment to consider the object $\gradt\ln p_\theta\left[\bf{X}_{0,T}|x_0\right]$.
The factor multiplying the return is commonly referred to as the Malliavin weight in the stochastic analysis literature,\cite{Warren2013} and corresponds to a particular case of the eligibility traces found in reinforcement learning\cite{Sutton1988,Sutton2018,Precup2000,Degris2012,Watkins1989}, which we denote as $y_\theta(T)=\gradt\ln p_\theta\left[\bf{X}_{0,T}|\bf{x}_0\right]$.
It can be rewritten by substituting the path probability,
\begin{align}
y_\theta(t'')-y_\theta(t')=\int_{t'}^{t''}dt \, \dot{y}_\theta(t),
\end{align}
where
\begin{align}\label{eq:malliavinweightprop}
\dot{y}_\theta(t) =\left[\G^{-1} \cdot \left(\dot{\bf{x}}(t) - \bf{F}_\theta(t)\right)\right] \cdot\left[ \G^{-1} \cdot \gradt \bf{F}_\theta(t)\right]
\end{align}
is the integrand of the Malliavin weight.
%The dependence on the path up to time $t$, and the state at time $t$, of $y_\theta(t)$ and $\dot{y}_\theta(t)$ are left implicit, respectively, writing
%\begin{align}\label{eq:return-Malliavin}
%	\gradt\DKL=-\left\langle R\left[x_{0,T}\right]y(T)\right\rangle_{p_\theta}.
%\end{align}
%Alternatively, expanding the return in equation Eq.~\ref{eq:return-Malliavin}, we may split the path probabilities into future and past components
%\begin{align}
%	&\gradt\DKL\nonumber\\
%	&=-\int_0^Tdt\left\langle r(\dot{x},x,t)\left(
%		y_\theta(t^+)+\left[y_\theta(T)-y_\theta(t^+)\right]\right)\right\rangle_{p_\theta}\nonumber\\
%	&=-\int_0^Tdt\left\langle r(\dot{x},x,t)y_\theta(t^+)\right\rangle_{p_\theta},
%\end{align}
%where we used $t^+$ as a shorthand for $t+\epsilon$ for some small positive epsilon, and used \eqref{eq:path-total-deriv} to remove the future contribution.
%
%The Malliavin weight must be split just after the time $t$ in order to use \eqref{eq:path-total-deriv} since the reward at time $t$ depends on the depends on the immediate future of $x$.

Were we to stop at \eqref{eq:unsupervised-gradient}, we would proceed to optimize a generative model (the diffusive dynamics with our parameterized force) of the trajectories using a score-function based approach, similar to standard unsupervised learning.
However, following the methods of reinforcement learning, we can use a combination of the Markovianity of the generative model and other variance reduction techniques to produce a gradient estimator which is much more efficient to estimate.
To begin with, we can simplify \eqref{eq:unsupervised-gradient} by noting that due to Markovianity, the Malliavin weight only correlates with the return in the future, and we can rewrite the gradient as 
\begin{eqnarray}\label{eq:mcr}
	\gradt\DKL &=&-\left\langle\int_0^{T}dt R\left[\bf{X}_{t^-,T}\right]\dot{y}_\theta(t)\right\rangle_{p_\theta} \nonumber \\
	&=& \chi_\mathrm{MCR}(\theta,T),
\end{eqnarray}
where we used $t^-$ as a shorthand for $t-\epsilon$ for some small positive $\epsilon$. We refer to the optimization of the modified dynamics using this formulation of the gradient as $\chi_\mathrm{MCR}$, as it is analogous to the Monte-Carlo returns (MCR), or REINFORCE\cite{Williams1987,Williams1992} policy gradient algorithm in reinforcement learning. In the long observation time limit, employing this gradient in stochastic optimization reduces to previous variational Monte Carlo procedures.\cite{Das2019}

%This is a perfectly good point to discretize and construct an optimization algorithm, analogous to some recently used in the literature \cite{Das2019}.
This estimator of the gradient is non-optimal for two reasons. First, it requires evaluation of a two time correlation function. In steady state, stationarity can be invoked to eliminate one of those integrals, however under finite time conditioning this simplification is not possible. Second, it has a high variance and requires significant averaging to converge accurate gradients. This is because both the Malliavin weight and the return undergo a random walk with linearly increasing variance\cite{Warren2013}.
Building on the analogies with the reinforcement learning formalism we define a value function as a path average of the return,
\begin{eqnarray}
	V(\bf{x},t)=\left\langle R\left[\bf{X}_{t,T}\right]\right\rangle_{p_\theta,\bf{x}}.
\end{eqnarray}
conditioned on starting at the position and time, $\bf{x}_{t}=\bf{x}$. Introduced into the gradients of $\DKL$ in distinct ways, the value functions can be used to tame both problems of the naive MCR gradient estimate. 

First, we introduce a value function as a baseline that depends only on the state at the time $t$ in order to reduce the variance of the gradient.  We note that $\dot{y}_\theta(t)$ is linear in the noise and thus averages to zero when multiplied by a function of the state at or before $t$.
Defining a temporal difference error
\begin{align}
	&\delta\left[\bf{X}_{t^-,T},t \right] =R\left[\bf{X}_{t^-,T}\right]-V\left( \bf{x}_{t},t\right) ,
\end{align}
we write the dynamical gradient as
\begin{eqnarray}\label{eq:mcvb-grad}
	\gradt\DKL&=&-\left\langle\int_0^{T}dt \delta\left[\bf{X}_{t^-,T},t \right]\dot{y}_\theta(t)\right\rangle_{p_\theta} \nonumber \\
	&=& \chi_\mathrm{MCVB}(\theta,T)
\end{eqnarray}
where we have formally subtracted zero. We refer to this gradient estimator as $\chi_\mathrm{MCVB}$, for Monte Carlo Value Baseline (MCVB)\cite{Sutton2018}. The subtraction of the state point dependent value function reduces the variance of the gradient by accounting for the mean uncorrelated part of each return between $t^-$ and $T$ with $\dot{y}_\theta(t)$, focusing on how this return differs from the average behaviour encoded by the value function.

Second, we introduce a value function that encodes an estimate of the return in the future in order to further reduce the variance and also the complications associated with estimating the two-time correlation function. We can replace part of the return by a value function that is conditioned at some $\tau$, such that $t^-<\tau < T$,  
\begin{align}
\left\langle  R\left[\bf{X}_{t^-,T}\right] \dot{y}_\theta(t) \right \rangle =& \left\langle  V\left( \bf{x}_{t+\tau},t+\tau\right)  \dot{y}_\theta(t) \right \rangle \nonumber \\
& +\left\langle  R\left[\bf{X}_{t^-,t+\tau}\right]   \dot{y}_\theta(t) \right \rangle
\end{align}
where we set the value function to zero for $V(\bf{x},t)$ with $t>T$. Combining this value function form of the kernel of the gradient with the value baseline, we define another temporal difference error
\begin{align}
	&\delta'\left[\bf{X}_{t^-,t+\tau},t\right]\\
	&=V\left( \bf{x}_{t+\tau},t+\tau\right) +R\left[\bf{X}_{t^-,t+\tau}\right]-V\left( \bf{x}_{t},t\right) ,\nonumber
\end{align}
and we arrive at a distinct formulation of the gradient
\begin{eqnarray}\label{eq:ac-grad}
	\gradt\DKL&=&-\left\langle\int_0^{T}dt \, \delta'\left[\bf{X}_{t^-,t+\tau}, t \right]\dot{y}_\theta(t)\right\rangle_{p_\theta} \nonumber \\
	&=&\chi_\mathrm{AC}(\theta,T)
\end{eqnarray}
which we denote $\chi_\mathrm{AC}(\theta,T)$ for actor-critic gradient (AC) estimator, for the analogous algorithm in reinforcement learning.\cite{Sutton2018,Haarnoja2018} 
Here the value function is seen as criticizing the transitions generated by the dynamics, i.e. the actor. 
Variance reduction of gradient estimates is therefore achieved by replacing potentially noisy return samples with the average behaviour expected in the future of the $\bf{x}_{t+\tau}$ state.
In Sec.~\ref{Sec:FTE}, we will compare the accuracy and statistical efficiency of these three gradient estimators: MCR, MCVB, and AC. 
Before that we discuss how the value functions are simultaneously parametrized and learnt along side the modified force.

%Finally, we may introduce a baseline function which depends only on the state at the time $t$ by noting that, in a discretized representation, the Malliavin weight integrated over the state at the next timestep present in $\dot{x}$ provides zero: in other words, $\dot{y}\theta(t)$ is linear in the noise at time $t$ and thus averages to zero when multiplied by a function of the state at or before $t$.
%Thus, using the value function as the baseline and defining the temporal difference error
%\begin{align}
%	&\delta\left[x_{t^-,t+\tau},t^-\right]\nonumber\\
%	&=V\left[x(t+\tau),t+\tau\right]+R\left[x_{t^-,t+\tau}\right]-V\left[x(t),t\right],
%\end{align}
%we may write the dynamical gradient as
%\begin{eqnarray}
%	\gradt\DKL=-\left\langle\int_0^{T}dt \, \delta'\left[x_{t^-,t+\tau},t\right]\dot{y}_\theta(t)\right\rangle_{p_\theta}.
%\end{eqnarray}

\subsection{Parametrizing value functions}
While the gradient expressions are exact and the use of value functions expected to facilitate their convergence, using them requires knowledge of the exact value function for the modified dynamics, a formidable task in complex problems. In order to make their use tractable, we optimize a representation of the value function in addition to the modified force. Specifically, we introduce a parametrization of the value function denoted $V_\psi$.
To optimize this approximation we note that the value functions satisfy a self-consistency equation called the Bellman equation\cite{baird1999gradient}
\begin{align}\label{eq:bellman}
	V(\bf{x},t)=\left\langle V\left( \bf{x}_{t+\tau},t+\tau\right) +R\left[\bf{X}_{t,t+\tau}\right]\right\rangle_{p_\theta,\bf{x}},
\end{align}
which has a unique solution for a given dynamics and return (as defined by the tilting observable and the dynamics via \eqref{eq:returnfull}).
We aim to minimize the error in this equation, thus optimizing our parametrized value towards this unique solution.
Our approach is to minimize the squared difference between the two sides of \eqref{eq:bellman} with the true value function replaced by the parametrized value function, and apply gradient descent to it.
Such an approach is the subject of gradient temporal difference methods \cite{Sutton2009,Maei2009,Maei2011}, but produces a gradient estimate which is difficult to evaluate, containing products of expectations which require independent samples. A part of the resultant gradient is however simpler to compute. We derive it by
substituting only the right hand side of \eqref{eq:bellman} with our parametrized value function to provide a fixed target for the left and defining a corresponding error function based on the squared difference. 
To construct a loss, we integrate these errors along each trajectory, and average them over the trajectory ensemble. 
This results in a loss function $L(\psi,\psi_i)$, that we take as a function of two weights, $\psi$ and $\psi_i$,
\begin{align}
	&L(\psi,\psi_i)=\nonumber\\
	&\frac{1}{2}\Biggl\langle \int_0^Tdt \Big \{ \left\langle V_{\psi_i}\left( \bf{x}_{t+\tau},t+\tau\right) +R\left[ \bf{X}_{t,t+\tau}\right] \right\rangle _{p_\theta,\bf{x}} \nonumber\\
	&-V_\psi\left( \bf{x}_{t},t\right) \Big \}^2 \Biggr\rangle _{p_\theta} ,
\end{align}
where the weight $\psi_i$ is the weights after update $i$, used to provide the fixed target estimate towards which we want to move the functional of $\psi$.
The derivative is then taken with respect to $\psi$, before setting $\psi = \psi_i$ to find the gradient of this loss for the current parameters.
Such an approach is referred to as semi-gradient in the reinforcement learning literature,\cite{Sutton2018} used to achieve the majority of state-of-the-art reinforcement learning results, and proves stable provided the data used to estimate the gradient is sampled using a dynamics which is close to $p_\theta$ as we intend to do.
As mentioned above, alternative methods which additionally consider the variation of the target with $\psi$ can be found in the RL literature, allowing for the use of data sampled from an alternative dynamics, utilized via importance sampling.\cite{Sutton2009,Maei2009,Maei2011}

Writing an approximate temporal difference for the value function parametrization, within MCVB
\begin{align}\label{eq:mcvb-td}
	&\delta_\psi\left[\bf{X}_{t^-,T},t\right]=R\left[\bf{X}_{t^-,T}\right]-V_\psi\left( \bf{x}_{t},t\right) ,
\end{align}
or for AC
\begin{align}\label{eq:ac-td}
	&\delta'_\psi\left[\bf{X}_{t^-,t+\tau},t\right]\nonumber\\
	&=V_\psi\left( \bf{x}_{t+\tau},t+\tau\right) +R\left[\bf{X}_{t^-,t+\tau}\right]-V_\psi\left( \bf{x}_{t},t\right) ,
\end{align}
we have gradients of the form
\begin{align}\label{eq:vgrad}
	&\left.\gradp L(\psi,\psi_i)\right|_{\psi=\psi_i}\nonumber\\
	&=-\Biggl\langle\int_0^T dt\; \delta_{\psi_i}\left[\bf{X}_{t^-,T},t\right] \left.\gradp V_\psi\left( \bf{x}_{t},t\right) \right|_{\psi=\psi_i}\Biggr\rangle_{p_\theta} ,
\end{align}
for the loss function from the value function parametrization, where for the AC algorithm $\delta_{\psi_i}$ is replaced with $\delta_{\psi_i}'$. Given this value function approximation, we can approximate the gradient of the KL divergence by replacing the exact temporal difference with these approximate temporal differences. We then use the same trajectories to estimate the force and value function gradients and simultaneously learn both. For the MCVB algorithm, an approximate value function does not bias the gradients as the future return that correlates with the Malliavin weight stays intact and the expectation of the Malliavin weight is identically 0. However, for the AC algorithm, an approximate value function can introduce a bias into gradients as it replaces the average of the future return, which it may not accurately represent. 

Employing gradients with or without value functions, we can construct a stochastic descent algorithm to optimize the modified forces which can be used to estimate the likelihoods of rare events and the trajectories by which they emerge. The algorithms require the evaluation of the forces, value function, their parametric gradients and noises over the course of simulating trajectories. Ensembles of trajectories can then be used to construct an empirical estimate of the gradient via computing the Malliavin weights, returns, and the temporal difference.
These empirical estimates then iterate the two weights with respective learning rates $\alpha^\theta$ and $\alpha^\psi$ for the force and value function respectively. The resultant algorithm is outlined in pseudocode below in Alg.~\ref{finite_time_algo}. Detailed versions of the individual algorithms with computationally efficient on-the-fly implementations for simulating trajectories with discrete timesteps are presented in Appendix ~\ref{sec:discrete-timestep}.

\begin{algorithm}[H]
	\caption{Gradient optimization using finite time trajectories}\label{finite_time_algo}
	\begin{algorithmic}[1]
		\State \textbf{inputs} dynamical approximation $F_\theta(\bf{x},t)$, value approximation $V_\psi(\bf{x},t)$
		\State \textbf{parameters} learning rates $\alpha^\theta$, $\alpha^\psi$; total optimization steps $I$; trajectory length $T$ consisting of $J$ timesteps of duration $\Dt$ each; number of trajectories $N$
		\State \textbf{initialize} choose initial weights $\theta$ and $\psi$, define iteration variables $i$ and $j$, force and value function gradients $\delta_{P}$, $\delta_V$, temporal difference $\delta$ (can be $R\left[ \bf{X}_{t^{-},T}\right]$ or $\delta_{\psi}\left[ \bf{X}_{t^{-},T},t\right]$ or $\delta_{\psi}^{'}\left[ \bf{X}_{t^{-},t+\tau},t\right]$ for MCR/MCVB/AC)
		\State $i\gets0$
		\Repeat
		\State Using chosen method to generate trajectories $\bf{X}_{0,T}$ with configurations, times and temporal differences denoted by $\bf{x}_{j},t_{j}$ and $\delta_{j}$ respectively.
		\State $j\gets0$
		\State $\delta_P\gets0$
		\State $\delta_V\gets0$
		\Repeat
		\State $\delta_P\gets\delta_P+\delta_{j}\dot{y}_{\theta}(t_{j})\Dt$
		\State $\delta_V\gets\delta_V+\delta_j\gradp V_\psi(\bf{x}_j,t_j)\Dt$
		\State $j\gets j+1$
		\Until{$j=J$}
		\State average $\delta_{P}$,$\delta_{V}$ over $N$ trajectories to get $\overline{\delta}_{P}$, $\overline{\delta}_{V}$
		\State $\theta\gets\theta+\alpha^\theta\overline{\delta}_P$
		\State $\psi\gets\psi+\alpha^\psi\overline{\delta}_V$
		\State $i\gets i+1$
		\Until{$i=I$}
	\end{algorithmic}
\end{algorithm}

\section{Rare fluctuations in finite time}\label{Sec:FTE}
We have used the algorithms discussed above to examine rare fluctuations of trajectories of fixed duration, starting from a fixed point in configuration space. The specific observable we have investigated is an indicator function for reaching a desired region, $\Gamma$, in configuration space,
$O[\bf{X}_{0,T}]=h_{\Gamma}[\bf{x}_{T}]$, where
\begin{equation*}
    h_{\Gamma}[\bf{x}_{T}]=\left\{
	\begin{array}{l@{\qquad}l}
		1  & \bf{x}_{T}\in\Gamma \\
		0 & \mathrm{otherwise}
	\end{array}\right. \, ,
\end{equation*}
at the final time $T$. Rare trajectories reaching a target basin in configuration space are often of interest as transition paths for reactive events, and significant development has been undertaken to efficiently generate them.\cite{Bolhuis2002,invernizzi2020unified,khoo2019solving,li2019computing,rotskoff2020active} Computing optimal drift forces for generating these rare trajectories enables the study of reactive dynamics in a direct manner. We expect these algorithms to find use in the study of diffusive dynamics where Monte Carlo approaches have difficulty sampling.\cite{gingrich2015preserving,grunwald2008precision,guttenberg2012steered,stoltz2007path} Further, as the modified force is used with the original noise from the SDE, we have access to the full reactive trajectory ensemble allowing the interrogation of the statistics of the reactive events in a way that other direct path methods like nudged elastic band and zero temperature string methods do not, as they represent only the dominant path.\cite{henkelman2000climbing,henkelman2000improved,henkelman2001methods,weinan2002string} As a consequence, we expect out method will find use when there is a large path space entropy. 

\begin{figure*}[t!]
	\includegraphics[width=1\linewidth]{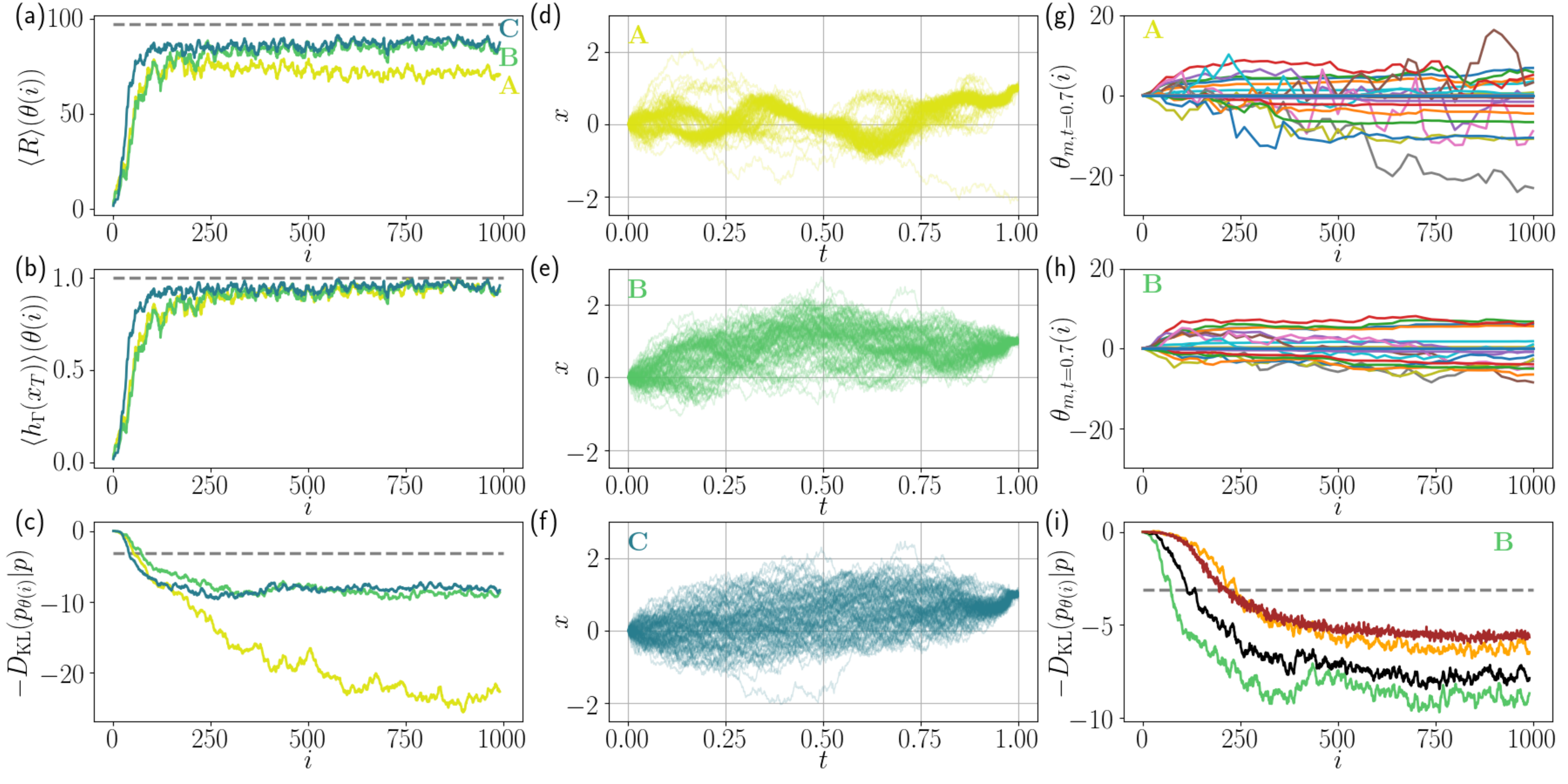}
	\caption{\label{fig:bridge} \textbf{Softened Brownian bridges:} (Left column) Smoothened learning curves showing running estimates of the CGF (a), average value of the indicator observable with the optimized dynamics (b), and the average cost function (c), as functions of optimization steps $i$, with the MCR(`A', yellow), MCVB(`B', green) and AC(`C',blue) algorithms. The horizontal dashed grey lines denote the numerically exact values. (Middle column) 100 trajectories obtained with the final converged dynamics from the three different algorithms but with the same noise history.(Right column) (g) and (h) show the smoothened convergence of a time-slice of the force parameters, as a function of optimization steps $i$, in the absence (MCR) and presence (MCVB) of a value function. (i) shows the convergence of the KL divergence cost with finer basis sets optimized with the MCVB algorithm. Green ($31x\times21t$), black ($31x\times41t$), orange ($31x\times81t$)  and brown ($41x\times201t$) curves show that in the increasing basis limit, the cost-function estimate approaches the value expected from the numerically exact CGF.}
	\label{fig1}
\end{figure*}

The CGF for an indicator variable is given by
\begin{equation}
    \lambda(s,T)=\ln\left\langle e^{-sh_{\Gamma}[\bf{x}_{T}]}\right\rangle_p
\end{equation}
as an average in the original reference dynamics. From Eq.~(\ref{eq:bound}), the KL divergence being nonnegative implies the average return is bounded above by the value of the CGF $\lambda(s,T)$. The bound can be saturated only by the unique optimal drift force. 
We compare the value of the optimized return to numerically exact estimates of the CGF given as
\begin{align}\label{eq:cgfform}
    \lambda(s,T)=\ln\left \{ 1+(e^{-s}-1)\int_{\Gamma}d\bf{x}\;\rho(\bf{x},T)\right \},
\end{align}
where the definition of the indicator function and the final time distribution $\rho(\bf{x},T)$ evolved from a specific initial condition has been used. This form demonstrates the statistics of a single-time indicator observable is described solely by its mean,
\begin{align}\label{eq:fprobab}
    \langle h_{\Gamma}\rangle_p =\int_{\Gamma}d\bf{x}\;\rho(\bf{x},T) \, .
\end{align}
For a rare fluctuation such that $\langle h_{\Gamma}\rangle_p<0.5$, this form indicates that there are two distinct regimes in the biased ensemble with $s<0$. For a small magnitude of the bias, the indicator function stays close to the unbiased value. Below a critical value of $s^{*}=-\ln[\langle h_{\Gamma}\rangle_p/(1-\langle h_{\Gamma}\rangle_p)]$ the indicator crosses over to being close to 1. For all of our calculations, we choose a fixed value of $s$ estimated to be smaller then the threshold. With this value of $s$, we compute the right side of Eq.~(\ref{eq:cgfform}) using an eigen-expansion of the propagator of the Fokker-Planck equation of the original dynamics, and compare with the value of the average return from the gradient descent algorithms having the same value of $s$. Details of this calculation and comparison to an approximate Kramers escape rate are in Appendix \ref{sec:altcgf}.

\subsection{Softened Brownian bridges}
The first example we consider is a softened version of a so-called Brownian bridge,\cite{revuz2013continuous,Majumdar2015} in which a one-dimensional Brownian motion starting from the origin is biased to end near a particular point. 
The reference dynamics is simply given by free diffusion, 
\begin{equation}
dx=\sqrt{2}d W
\end{equation}
where comparing to Eq.~\ref{eq:sde} we have $G=\sqrt{2}$. We consider the target well, $\Gamma(x)$, to be defined as $\{1-\epsilon \le x \le1+\epsilon \}$ with $\epsilon=0.1$. The dynamics is simulated with a discrete timestep of 0.001. We use a tilting parameter $s=-100$ to bias the original ensemble towards higher occurrence of the rare event.

We optimize a force and value function parametrized by linear combinations of Gaussian distributions with fixed variance and mean. Given a set of means $\{(x_m,t_m)\}_{m=0}^M$ and variances $\{\sigma_m\}_{m=0}^M$, the force and value function of a position $x$ at time $t$ are given by the coefficients $\{\theta_m\}_{m=0}^M$ and $\{\psi_m\}_{m=0}^M$ as
\begin{eqnarray}
	F_\theta(x,t)=F(x)+\sum_{m=0}^M\theta_m e^{-\frac{(x-x_m)^2+(t-t_m)^2}{2\sigma_m}} \nonumber \\
	V_\psi(x,t)=\sum_{m=0}^M\psi_m e^{-\frac{(x-x_m)^2+(t-t_m)^2}{2\sigma_m}} , 
\end{eqnarray}
where initially the basis sets are a grid of $31\times21$ Gaussians in the $x$-$t$ space. The Gaussians in time are spaced uniformly between $t\in[0,T)$, with standard deviations equal to half the grid-spacing. A third of the Gaussians in space are placed between $x\in[-4,-0.5]$, a third in $x\in(-0.5,1.5)$ and a third in $x\in[1.5,5]$. These three families of  Gaussians each have standard deviations half of the corresponding grid spacings. We initialize all $\theta_{m}=\psi_{m}=0$.

We consider the performance of the three algorithms differing in the gradient used to optimize them. These include an algorithm that uses no value function (MCR), one that uses a value baseline (MCVB), and one that uses a value function for future returns with $\tau=0.1$ (AC). We evaluate the efficiency of the algorithms by comparing learning curves, convergence with respect to basis, and properties of the learnt dynamics, shown in Fig~\ref{fig1}. All figures comparing different algorithms use the same noise history and the same amount of statistics, such that the differences are solely ascribed to the learned dynamics. The MCR algorithm uses a learning rate of $\alpha^{\theta}=0.4$. The MCVB algorithm learning rates $\alpha^{\theta}=0.4,\alpha^{\psi}=50$, and the AC algorithm learning rates $\alpha^{\theta}=1,\alpha^{\psi}=0.05$. 

In Figs.~\ref{fig1}(a-c), we show learning curves for the total return, the average of the indicator observable, and the KL divergence, generated with 12 trajectories at each optimization step for each of the three algorithms. We have compared the results obtained with this finite basis to the numerically exact value of the optimal return and the corresponding observable average and KL divergence, obtained from Eq.~\ref{eq:cgfform} where for free diffusion the distribution is known. We find that while all three algorithms quickly achieve a dynamics which mostly fulfills the indicator function conditioning, the MCR algorithm struggles to optimize the KL divergence cost, while the MCVB and AC algorithm achieve converged values efficienctly. As expected, each algorithm provides a variational estimate to the CGF with the MCVB and AC outperforming MCR. Trajectories with the final learned dynamics for the three algorithms are plotted in Fig.~\ref{fig1}(d-f). The MCR algorithm finds forces that constrain the bridge trajectories too excessively, which results in the suboptimal estimate of the KL divergence. The AC trajectories are closest to the optimal bridge trajectories\cite{Majumdar2015} while the MCVB trajectories lie in between. The main reason for the difference in performance in the three algorithms is the resultant suppression in the statistical errors in the gradient estimate. This is illustrated in Figs.~\ref{fig1}(g-h) where the convergence of the gradients of the 31 Gaussian coefficients at a time slice of $t=0.7$ is shown for both MCR and MCVB. Since the $\alpha^{\theta}$ learning rate is same in both algorithms, the large suppression of fluctuations in the MCVB learning curves results from a more statistically converged gradient estimate using a value function. This suppression of gradient errors at limited statistics in the MCVB and AC algorithms is directly illustrated in Appendix ~\ref{sec:error}.

We have studied the convergence of the KL divergence estimate towards the optimal value extracted from the numerically exact CGF, using the MCVB algorithm with an increasing position and time basis. We increased the number of time Gaussians, from 21 to 41 to 81, to observe the KL divergence cost shrinking as the finer grained force can better support the singular indicator function condition at the end of the trajectory. We also ran the optimization with a much bigger basis of $41x\times201 t$ Gaussians, and used 248 trajectories at every optimization step and learning rates $\alpha^{\theta}=5,\alpha^{\psi}=1000$. The Gaussians in $x$ have standard deviations equal to half the grid spacing, while the Gaussians in $t$ have standard deviations equal to a third of the grid spacing. While the estimate increased, in this particular problem, obtaining the numerically exact KL divergence would require use of still finer-grained Gaussians in space and time in order to represent the singularities of the edges of the target region and of the last timestep.

\subsection{Barrier crossing with multiple reaction pathways}\label{sec:2dL}

\begin{figure*}[t!]
	\includegraphics[width=1\linewidth]{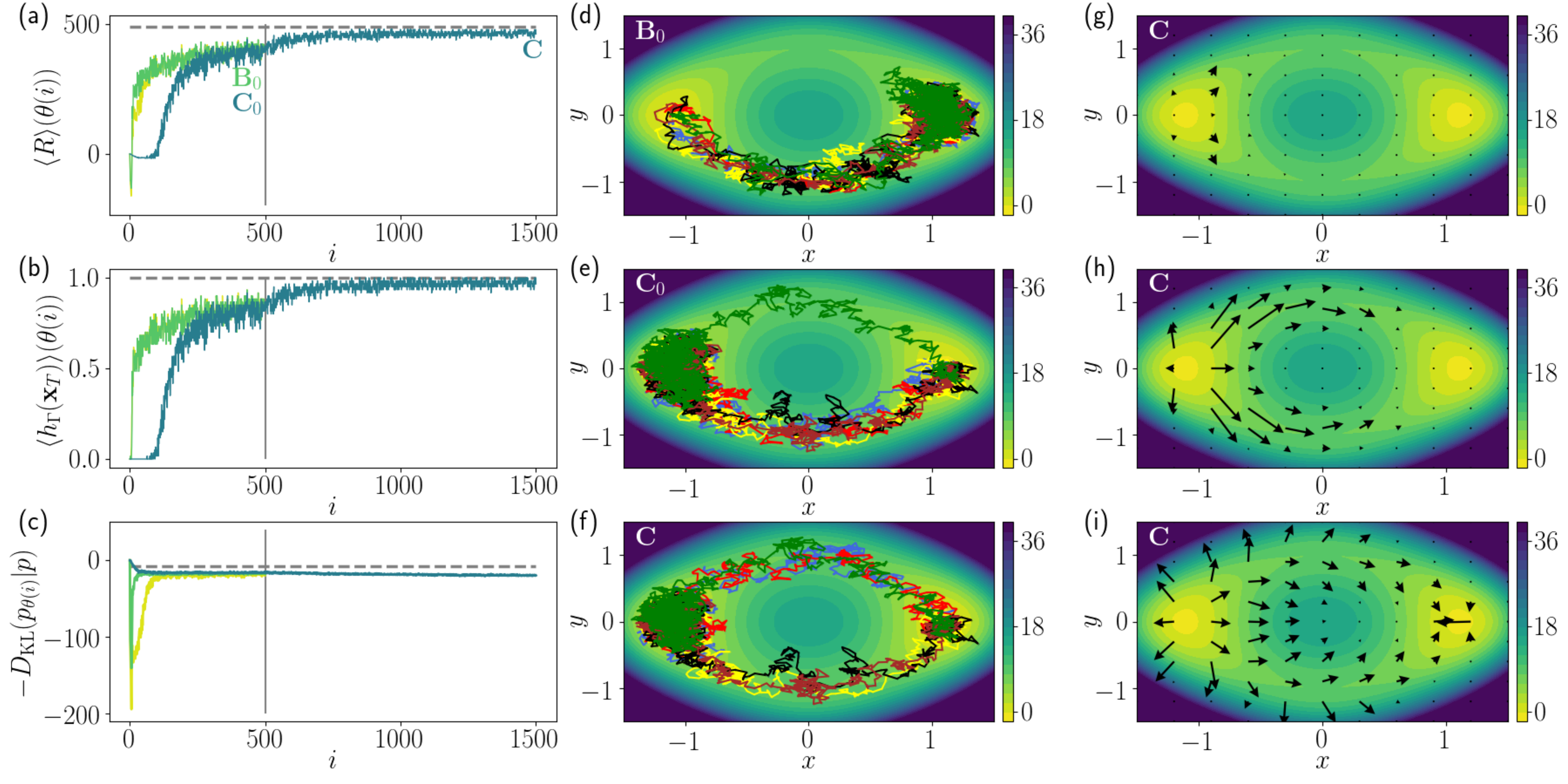}
	\caption{\label{fig:tube} \textbf{Multiple reaction pathways:} (Left column) Smoothened learning curves showing running estimates of the CGF (a), average value of the indicator observable with the optimized dynamics (b), and the average cost function (c), as functions of optimization steps $i$, with the MCR(yellow), MCVB(green) and AC(blue) algorithms. The vertical grey lines denote the end of initialization and beginning of optimization run. The horizontal dashed grey lines denote the numerically exact values. The parameter values from the end of the initialization with MCVB and AC have been called $B_{0}$ and $C_{0}$ respectively. The forces at the end of optimization with AC is called $C$. (Middle column) 6 representative trajectories obtained with the forces $B_{0}$ (panel d), $C_{0}$ (panel e), and $C$ (panel f). (Right column) Two-dimensional vectorial representation of the spatially dependent forces as a function of time, at $t=1$ (g), $t=1.3$ (h) and $t=1.5$ (i), obtained from the converged parameters at $C$.}
\end{figure*}

We now investigate the ability of the three algorithms to find the optimal dynamics in two-dimensional barrier-crossing problems, the first involving a potential allowing for multiple reaction pathways.
The two-dimensional potential $U(\bf{x})$ we consider\footnote{The potential we use is $U(x,y)=4/3 [4(1-x^{2}-y^{2})^{2}+2(x^{2}-2)^{2}+((x+y)^{2}-1)^{2}-((x-y)^{2}-1)^{2}-2]$}
 has two minima and two degenerate reaction pathways involving the upper and lower halves of the $\bf{x}=(x,y)$ plane as illustrated in Fig.~\ref{fig:tube}. Barrier-crossing from one well to another is a rare event occurring with one randomly chosen pathway.\cite{dellago1998transition} Without prior knowledge of the possibility of multiple reaction paths, path sampling algorithms typically need special techniques to discover them.\cite{fujisaki2010onsager} We use our reinforcement learning algorithms to compute an optimal force $F_\theta(\bf{x},t)$ that reproduces unbiased and uncorrelated reaction paths.

The reference equation of motion we consider  is
\begin{equation}
\label{Eq:2dL}
d{\mathbf{x}}=-\nabla U(\mathbf{x})+\sqrt{2}d\boldsymbol{W}
\end{equation}
where the matrix $\G$ is proportional to the identity. We use a discretization timestep of 0.001. The trajectories start from the minimum of the left well, at $(x,y)=(-1.11,0)$, and are allowed to run for a duration of $T=1.5$ and checked for reaching the right target well defined as $x>0,U(x,y)<0$. This small region centered around (1.11,0) is used as $\Gamma$ for defining the indicator function observable. The value of $T$ has been chosen to be slightly greater than the typical transition path timescale, such that the optimized force should reproduce trajectories that follow the natural steady-state fluctuations of the system. As long as the choice of $T$ is arbitrarily larger than the typical transition path timescale, the optimally generated trajectories will represent unbiased reactive transitions, with additional times being spent in the initial or final metastable states.\cite{delarue2017ab} In the absence of an approximate transition path time estimate, the optimization can be performed over a range of $T$ increasing by orders of magnitude till one enters the regime where side-side correlation functions for the dynamics of barrier crossing behave linearly.\cite{dellago1998transition} We use a value of $s=-500$ to obtain the CGF. The force and the value function are approximated again as a grid of Gaussians with optimizable coefficients, a simple generalization of the one-dimensional Brownian bridge. 

The duration of the trajectories we consider, $T$, is much smaller than the typical first passage time for the rare fluctuation we are interested in studying. As such, a general complication arises in initializing our algorithms in that in the absence of a modified force, few trajectories satisfy the indicator function condition. Consequently, the gradients for updating the modified forces are generally very small and noisy. In order to initialize our learning process, we start with a softened version of the indicator function of the form
\begin{equation}
\label{eq:quad}
\tilde{h}[\bf{x}_{T}]=-[(x_{T}-x_{f})^{2}+(y_{T}-y_{f})^{2}]
\end{equation}
which is quadratic, and non-vanishing across the full domain. After optimizing the return with this observable, we obtain a force that can surpass the barrier, and  the optimization with the sharp indicator function observable can begin. This technique of breaking down the optimization of the return into two segments prioritizing each of the two terms of the return is analogous to curriculum learning in reinforcement learning.\cite{bengio2009curriculum} In many-body systems, the quadratic metric can be defined only in the space of the order parameter that distinguishes the initial and product states. For our multi-channel problem, we initialize learning with $(x_{f},y_{f})=(1.11,0)$ in the softened indicator, which is the minimum of the target well.  Our approach consists of comparing the performance of the three algorithms MCR, MCVB and AC in the initialization with the quadratic observable, and then using the AC algorithm to optimize the return with the indicator function observable. 

Figures~\ref{fig:tube}(a-c) demonstrate the learning curves for the full return, the average of the indicator function and the KL divergence cost. The three initializations each use 60 trajectories at every optimization step. The basis functions for the force and value function used are a grid of $21\times21\times41$ Gaussians in the $\bf{x}-t$ space for each component independently. The Gaussians are placed uniformly on the time axis $t\in[0,T)$, while the position Gaussians are distributed uniformly between $x\in[-1.5,1.5]$ and $y\in[-1.5,1.5]$. The learning rates used in the initialization are $\alpha^{\theta}=1$ for MCR, $\alpha^{\theta}=1,\alpha^{\psi}=0.5$ for MCVB and $\alpha^{\theta}=1,\alpha^{\psi}=0.5, \tau =0.001$ for AC, and the learning rate for the final optimization is $\alpha^{\theta}=0.2,\alpha^{\psi}=0.08, \tau=0.1$ in the AC algorithm. In the learning curves, we compare the convergence of the return with numerically exact values obtained by computing the RHS in Eq.~\ref{eq:cgfform} with a spectral expansion using a Discrete Variable Representation basis.\cite{szalay1993discrete} We see  that all three algorithms quickly find forces that satisfy the conditioning, but the KL divergence cost is optimized best by the AC algorithm. While each affords a similar variational estimate after the initial optimization, we find qualitative differences in the family of barrier-crossing trajectories obtained from the MCR/MCVB and from the AC algorithm. 

\begin{figure*}[t!]
	\includegraphics[width=1\linewidth]{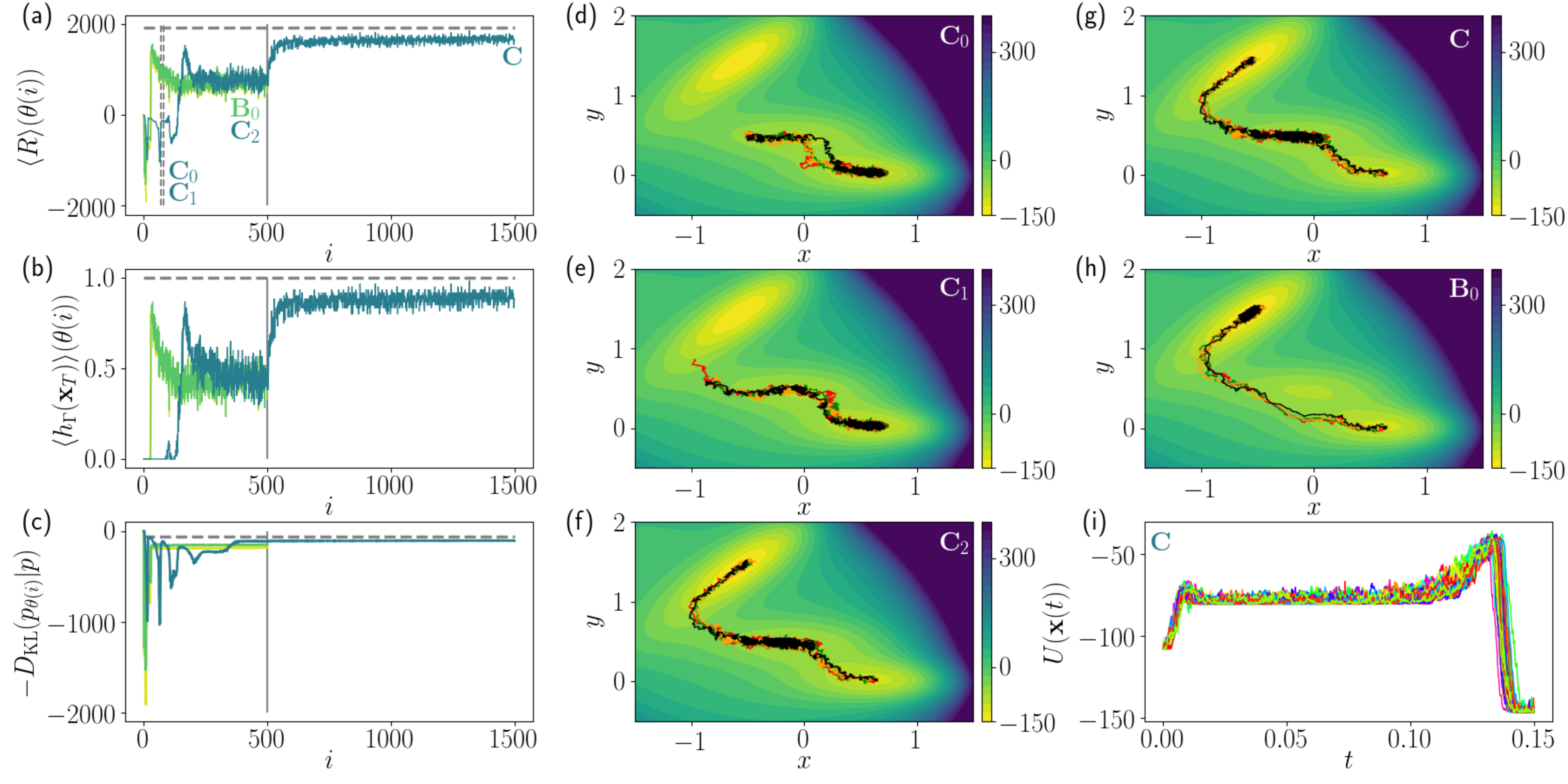}
	\caption{\label{fig:mbpot} \textbf{M\"uller-Brown potential:} (Left column) Smoothened learning curves showing running estimates of the CGF (a), average value of the indicator observable with the optimized dynamics (b), and the average cost function (c), as functions of optimization steps $i$, with the MCR(yellow), MCVB(green) and AC(blue) algorithms. The vertical grey lines denote the end of initialization and beginning of optimization run. The horizontal dashed grey lines denote the approximate values from a Kramer's escape rate approximation. On the AC learning curve in (a), the parameter values at $i=70$ and $i=80$ (the vertical dashed lines) have been called $C_{0}$ and $C_{1}$ respectively. The values at the end of initialization with MCVB and AC are called $B_{0}$ and $C_{2}$, and at the end of AC optimization are called $C$. (Middle column) 4 representative trajectories obtained with the forces $C_{0}$(d), $C_{1}$(e), and $C_{2}$(f). (Right column) 4 representative trajectories obtained with the forces $C$(g) and $B_{0}$(h). (i) Potential energy as a function of time for 100 representative trajectories driven with the force parameters $C$.}
\end{figure*}
Typical trajectories obtained with forces from the end of initialization with MCVB and AC, and at the end of optimization with AC, are shown in Figs.~\ref{fig:tube}(d-f). The force obtained from MCVB spontaneously breaks the symmetry in the potential and chooses one reaction path out of the two. This force solution is a local optimum in the MCR and MCVB algorithms, and it does not naturally relax to a symmetric force that would be representative of the degeneracy of the reaction paths. Trajectories from the AC algorithm spend significant amount of time exploring the initial well, such that the discovered forces recognize the presence of multiple pathways approximately. These forces are further refined during the second optimization, such that the reactive trajectories obtained at the end are restored to be almost fully symmetric like the natural barrier-crossing fluctuations of the system are expected to be. These symmetric two-dimensional forces obtained at the end of the AC optimization are plotted at three slices of time, in Figs.~\ref{fig:tube}(g-i). The forces grow in magnitude as a function of time and generally follow the contours of the underlying potential, and towards the end they gather support in unlikely parts of the potential. The ability of the AC algorithm to discover time-dependent forces that lead to exploration of multiple reaction pathways can prove valuable in uncovering reactive trajectories in systems where such degeneracies are not known \textit{a priori}. 

\subsection{Barrier crossing with a long lived intermediate}

Another difficult problem in the generation of transition paths and reactive trajectories typically comes from the presence of long-lived intermediates. In order to study the usefulness of our learning algorithms in this context, we consider as an example the dynamics on the so-called M\"uller-Brown potential.\cite{muller1979location}
This two-dimensional potential surface has been used extensively as a testing case for methods relying on the instantonic approximation for barrier-crossing trajectories.\cite{delarue2017ab,bonfanti2017methods} The potential is a sum of four Gaussians\footnote{
The potential takes the form, $U(x,y)=\protect\sum_i A_{i}\exp[a_{i}(x-\bar{x}_{i})^{2}+b_{i}(x-\bar{x}_{i})(y-\bar{y}_{i})+c_{i}(y-\bar{y}_{i})^{2}]$ where
$A=(-200,-100,-170,15)$, $a=(-1,-1,-6.5,0.7)$,  $b=(0,0,11,0.6)$, $c=(-10,-10,-6.5,0.7)$, $\bar{x}=(1,0,-0.5,-1)$, and $\bar{y}=(0,0.5,1.5,1)$.}, where three local minima are separated by two barriers as illustrated in Fig.~\ref{fig:mbpot}. We employed our algorithms to find forces that generate uncorrelated trajectories that cross both barriers, starting from a local minimum and ending in the global minimum, that are positioned on either side of the third metastable minimum. 

The system evolves with diffusive Langevin dynamics of the same form as Eq.~\ref{Eq:2dL} using a timestep of $0.0001$. We are interested in trajectories starting from $\bf{x}=(0.63,0.03)$ in the rightmost local minumum, and ending near the global minimum, centered around $\bf{x}=(-0.5,1.5)$, with the indicator function region $\Gamma$ being defined by $U(\bf{x})<145$).  The trajectories are chosen to be of a fixed duration of $T=0.15$, which is on the order of the expected total transition path timescale from Kramers' theory added to the expected relaxation time in the intermediate well.\cite{laleman2017transition,delarue2017ab} For initializing the forces we use a softened quadratic modification of the indicator, in Eq.~\ref{eq:quad}, with $s=-10000$, while we use a bias value of $s=-2000$ with the indicator observable to compute the CGF.  To represent the $x$ and $y$ components independently of the time-dependent optimal force and to represent the value function, we use a basis of Gaussians with optimizable coefficients placed on a $21\times21\times21$ grid in $\bf{x}-t$. The time Gaussians are placed uniformly between $t\in[0,T)$, while the space Gaussians placed uniformly between $x\in[-1.5,1.5]$ and $y\in[-0.5,2]$.

In Figs.~\ref{fig:mbpot}(a-c), we have compared the learning curves with MCR, MCVB and AC algorithms during initialization with the smooth indicator function in Eq.~\ref{eq:quad} and the AC algorithm for the final optimization of the full return with the sharp indicator function. Each algorithm uses 60 trajectories at every optimization step to estimate the gradient. The learning rates for the initialization are $\alpha^{\theta}=1$ for MCR, $\alpha^{\theta}=1,\alpha^{\psi}=1$ for MCVB, and $\alpha^{\theta}=0.5,\alpha^{\psi}=0.2,\tau=0.0001$ for AC, and the learning rates for the final optimization are $\alpha^{\theta}=0.1,\alpha^{\psi}=0.01,\tau=0.01$ for AC. The learning curves have been compared with approximately calculated values of the CGF and the KL div obtained with a Kramer's escape rate estimate along the Minimum Energy Path.\cite{henkelman2000climbing} 

We find that all three algorithms optimize the quadratic observable relatively quickly, but the AC algorithm performs the best at optimizing the KL divergence cost. In Figs.~\ref{fig:mbpot}(d-h), we illustrate a few uncorrelated trajectories generated with the modified forces at various stages of the initialization and optimization with the AC method and the end of the initialization with the MCVB method. We find that the forces with the AC algorithm are such that the trajectories discover and cross the two barriers and the metastable well between them one after another. At the end of the AC initialization, the trajectories have discovered the metastable well and have crossed both barriers to end in the target well. The AC algorithm by this stage of optimization has also moved the major part of the short trajectory from staying in the initial well to the metastable well. This feature is constant throughout the AC optimization, with only minor changes in the force being carried out inside the target end well. The force from the MCVB initialization, on the other hand, only generates trajectories that connect the initial and target well without relaxing significantly in the metastable well. This would be contrary to the instantonic relaxation mechanism in the system, as the stochastic action is minimized by the local relaxation in the metastable well. In Fig.~\ref{fig:mbpot}(i) we have plotted the potential energy as a function of time, for 100 uncorrelated barrier-crossing trajectories, which are driven by the final force from the AC algorithm. The trajectories cross the two barriers at roughly fixed times, and spend majority of the time in the metastable well. 

The comparison of the three algorithms illustrates the significant improvement of convergence performance of the MCVB and AC algorithm over the naive MCR approach afforded by value functions. For rare reactive events, we have found that the AC algorithm is suited best to find trajectories that explore  configuration space the most in search for the easier barriers to cross, and thus is closest in resembling the natural fluctuations of the system. The errors in the converged values of the CGF depend on the truncation of the force basis and statistical uncertainties. The MCVB and AC algorithms preserve the computational scaling of the MCR with the trajectory duration, and only change the prefactors of the scaling by a small fraction making them viable methods for applications to complex systems. The AC algorithm with a small $\tau$ will incur a systematic error in the gradients if the value approximation is not accurate, which goes away at an intermediate $\tau$ but at the expense of a larger memory cost that may slow down the algorithm without any change in the scaling. Nevertheless, it is possible to use these algorithms with useful combinations of hyperparameters to achieve efficient convergence with a small amount of averaging. The value functions obtained during the optimizations serve as dynamical equivalents of the committor function, in that they encode the expected value of the probability to reach the target well and the associated KL divergence cost, while starting from any point in configuration space at any point in time. Understanding these connections to reaction coordinate design is likely a fruitful future direction of research.

\section{Gradient optimization for infinite time dynamics}\label{sec:gradient-opt-infinite}

We now generalize the approach of the previous section to focus on the statistics of time-integrated quantities in the long time limit. While for finite time, the generalized Doob transform is time dependent, under mild assumptions in the long time limit the optimal  dynamics is time-homogeneous.\cite{Chetrite2014} As a consequence, the parametrization of the modified force and value function is simplified, and explicitly dependent only on the instantaneous configuration of the system. The generalization of the algorithms to this case consists of two main changes. First, we employ online learning, since there is no end to each trajectory. Second, a modified definition of return and value are required to avoid divergences in the infinite time limit.

We formulate the infinite time problem by adapting an approach in reinforcement learning based on time-averaged returns \cite{Marbach2003,Schwartz1993,Bertsekas1996,Tsitsiklis1999}.
Specifically, we consider the long-time average of the KL divergence of the trajectory ensemble.
Under assumptions of time-independence and ergodicity,
\begin{align}\label{eq:continuing_KL_divergence}
	\dKL
	&=\lim_{T\rightarrow\infty}\frac{1}{T}\DKL\nonumber\\
	&=-\langle r(\bf{x},\dot{\bf{x}})\rangle_{p_\theta}+\lambda(s),
\end{align}
the time average KL divergence reduces to an average over the steady state distribution of the instantaneous change of the return $r(\bf{x},\dot{\bf{x}})$. Above, we have defined a scaled CGF,
\begin{eqnarray}\label{scgf}
    \lambda(s)=\lim_{T\rightarrow\infty}\frac{1}{T}\ln Z(s,T),
\end{eqnarray}
that is finite as long as the cumulants of the time-integrated observable are time extensive. The reward, $r(\bf{x},\dot{\bf{x}})$, is  defined as
\begin{eqnarray}
r(\dot{\bf{x}},\bf{x}) &=& -sA[\bf{x}]- s\bf{B}[\bf{x}]\cdot \dot{\bf{x}}  \\
&&+ \frac{1}{2} \left \{ | \G^{-1}\cdot\left(\dot{\bf{x}} - \bf{F}_\theta\right)| ^2 -| \G^{-1}\cdot\left(\dot{\bf{x}} - \bf{F}\right)|^2 \right \} \nonumber 
\end{eqnarray}
and is time-independent and evaluatable within the steady state.  A gradient expression analogous to MCR can be derived straightforwardly.\cite{Das2019}

The previous definition of the value will diverge in the infinite time limit.
A simple modification to address this issue is to remove the average reward scaled by the length of the trajectory segment, defining a differential return
\begin{eqnarray}
	\Delta R\trajarg
	&=& R\trajarg-(t'-t)\langle r(\dot{\bf{x}},\bf{x}) \rangle_{p_\theta} 
\end{eqnarray}
and corresponding differential value function
\begin{eqnarray}
	V(\bf{x})=\lim_{T\rightarrow\infty}\left\langle \Delta R\left[\bf{X}_{0,T}\right]\right\rangle_{p_\theta,\bf{x}}.
\end{eqnarray}
which satisfies a modified Bellman equation
\begin{eqnarray}\label{eq:differential_bellman}
	V(\bf{x})=\left\langle V\left( \bf{x}_{\tau}\right) +\Delta R\left[\bf{X}_{0,\tau}\right]\right\rangle_{p_\theta,\bf{x}},
\end{eqnarray}
containing the differential return between states, rather than the standard return, and relating the value of states separated by a period of time $\tau$.

This modified Bellman equation can be simply rearranged to give an alternative equation for our time-averaged KL divergence
\begin{eqnarray}
    \dKL=
    &-&\frac{1}{\tau}\left\langle V\left( \bf{x}_{\tau}\right) +R\left[\bf{X}_{0,\tau}\right]-V(\bf{x})\right\rangle_{p_\theta,\bf{x}}\nonumber\\
    &+&\lambda(s),
\end{eqnarray}
which we note holds for all $\bf{x}$.
Differentiating the right side of this equation with respect to $\theta$ does not involve the gradient of the stationary state.
Therefore, taking the derivative and then averaging over the stationary state under $F_\theta$,\footnote{Taking this derivative results in gradients of the value function at $x$ and $x_T$ with respect to $\theta$, however, these cancel out when averaging over the stationary state.} we can write an estimate of the dynamical gradient as
\begin{eqnarray}\label{eq:differential-dynamics-loss}
\gradt\dKL
=-\frac{1}{\tau}\left\langle\delta\left[\bf{X}_{0,\tau'}\right]y_\theta(\tau)\right\rangle_{p_\theta},
\end{eqnarray}
where we have defined the differential temporal difference error
\begin{eqnarray}
	\delta\left[\bf{X}_{0,\tau'}\right]=V\left( \bf{x}_{\tau'}\right) +\Delta R\left[\bf{X}_{0,\tau'}\right]-V\left(\bf{x}_{0}\right),
\end{eqnarray}
reached after introducing an additional baseline in the form of $\tau'\langle r(\dot{\bf{x}},\bf{x}) \rangle_{p_\theta}$.
In this equation we have arrived at a gradient estimate which depends only on the gradient of the transition probabilities, contained in the Malliavin weights $y_\theta(\tau)$, and not the gradient of the stationary state itself.
This can thus be easily calculated during a simulation using the parametrized dynamics.

Note the period of time $\tau'$ over which the temporal difference is calculated is independent of the period of time $\tau$ over which the Malliavin weight is calculated, provided the former is longer.
The specific algorithm we consider involves taking the time $\tau$ small enough so that the Malliavin weight can be approximated by  $\tau \dot{y}_\theta[\bf{x}_{0}]$ which is possible due to the time homogeneous steady state we average within.
We thus calculate the estimate as
\begin{eqnarray}
\gradt\dKL
&=&-\left\langle\delta\left[\bf{X}_{0,\tau'}\right]\dot{y}_\theta(0)\right\rangle_{p_\theta} \nonumber \\
&=& \chi_\mathrm{AC}(\theta)
\end{eqnarray}
which we denote as the actor-critic gradient in the long time limit. 
In practice, we will take $\tau'=\Dt$, a single time-step in a numerical simulation.
A long time limit generalization of the MCVB gradient could be constructed similarly, but this is not considered here.

As in the finite time case, to construct this estimate we also need an approximation to the value function, $V_\psi(\bf{x})$.
Following a similar construction for the loss function as before, averaging the error over the stationary state, we estimate the gradient by which to update the value function parameters as
\begin{eqnarray}
\gradp L(\psi)
=-\left\langle\delta_\psi\left[\bf{X}_{0,\tau'}\right]\gradp V_\psi\left( \bf{x}_{0}\right) \right\rangle_{p_\theta},
\end{eqnarray}
with the approximate temporal difference 
\begin{eqnarray}
	\delta_\psi\left[\bf{X}_{0,\tau'}\right]=V_\psi\left( \bf{x}_{\tau'}\right) +\Delta R\left[\bf{X}_{0,\tau'}\right]-V_\psi\left( \bf{x}_{0}\right) ,
\end{eqnarray}
which also replaces the exact temporal difference in gradient estimates for the dynamics. Finally, we also have flexibility with our estimate of the scaled CGF. This can be done using a running average of the reward, 
\begin{eqnarray}
    \langle r \rangle_{p_{\theta_i}} = \langle r \rangle_{p_{\theta_{i-1}}} + \alpha_r ( \langle r \rangle_{p_{\theta_{i}}}- \langle r \rangle_{p_{\theta_{i-1}}})
\end{eqnarray}
where $\alpha_r$ is the learning rate and the subscript $p_{\theta_{i}}$ denotes the parameters from the $i$th iteration. Alternatively, a lower variance, higher bias estimate may be constructed by noting that we can rearrange Eq.~\ref{eq:differential_bellman} to find
\begin{eqnarray}
    \langle r \rangle_{p_{\theta_i}} =\langle r \rangle_{p_{\theta_{i-1}}} + \alpha_r \langle \delta_\psi[\bf{X}_{0,\tau'}] \rangle_{p_{\theta_i}},
\end{eqnarray}
 an alternative equation for the average. After discretization, an algorithm based on utilising single-transition estimates of these gradients is outlined in pseudocode below in Alg.~\ref{differential_actor_critic}.

\begin{algorithm}[H]
	\caption{KL regularized differential actor-critic}\label{differential_actor_critic}
	\begin{algorithmic}[1]
		\State \textbf{inputs} force approximation $\textbf{F}_\theta(\textbf{x})$, value approximation $V_\psi(\textbf{x})$
		\State \textbf{parameters} learning rates $\alpha^\theta_i$, $\alpha^\psi_i$, $\alpha^R_i$; total updates $N$
		\State \textbf{initialize} choose initial weights $\theta$ and $\psi$, initial average $\bar{r}$, define iteration variable $i$, individual error $\delta$
		\State $i\gets0$
		\Repeat
		\State Generate a transition from $\textbf{x}$ to $\textbf{x}'$ according to the dynamics given by $\textbf{F}_\theta(\textbf{x})$ and noise vector $\textbf{w}\sim\mathcal{N}(0,1)$
		\State $\dot{y}_\theta=\frac{\textbf{w}^T\mathbb{G}^{-1}\gradt \textbf{F}_\theta}{\sqrt{\Dt}}$
		\State $\delta\gets V_\psi(\textbf{x}')+r(\textbf{x},\textbf{x}')-\bar{r}-V_\psi(\textbf{x})$
		\State $\theta\gets\theta+\alpha^\theta_i\delta\dot{y}_\theta$
		\State $\psi\gets\psi+\alpha^\psi_i\delta\nabla_\psi V_\psi(\textbf{x})$
		\State $\bar{r}\gets\bar{r}+\alpha^R_i\delta$
		\State $i\gets i+1$
		\Until{$i=N$}
	\end{algorithmic}
\end{algorithm}

\section{Rare fluctuations in the long time limit}\label{sec:velocity-ring}
Here we apply our approach to study the statistics of time-integrated currents in the long time limit.   Persistent currents are the hallmark of a nonequilibrium system, and their fluctuations have been studied intensively \cite{derrida2007non,pietzonka2016universal,bodineau2004current,grandpre2018current}. Foundational results have been derived that constrain the symmetries of current fluctuations and relate their cumulants. For example, the fluctuation theorems dictate that the CGF satisfies a reflection symmetry about the driving force for the current, due to the microscopic reversibility of the underlying stochastic dynamics.\cite{lebowitz1999gallavotti,crooks1999entropy} A number of numerical approaches have been developed to evaluate the scaled cumulant generating function, an example of a large deviation function \cite{Touchette2009,Bolhuis2002,Cerou2007,Lecomte2007,giardina2006direct,lestang2018computing,TsobgniNyawo2016}. These functions provide information of the long time behavior of stochastic systems, and encode response relationships and stability. Within this context, our approach is similar to other controlled dynamics \cite{dolezal2019large,Nemoto2016,Oakes2020,Whitelam2020,Kappen2016,Ray2018,Ferre2018,Das2019} based means of evaluating large deviation functions in the continuum and can be used directly as we show below or in concert with Monte Carlo algorithms.

To study the accuracy and efficiency of the algorithm, we consider statistics of the velocity of a particle on a ring of length $L=2\pi$ with position $x$ moving in a periodic potential. The periodic potential has the form $U(x)=U_0\cos(x)$ with magnitude $U_0$, and is driven by a constant force $f$, such that
\begin{eqnarray}
    F(x)=-\frac{dU(x)}{dx}+f
\end{eqnarray}
is the total force for the particle on the ring. The observable we consider is the integrated current, $O[\bf{X}_{0,T}] =J[\bf{X}_{0,T}]$ given by
\begin{eqnarray}
J[\bf{X}_{0,T}] = \int_0^T dt \, \dot{x}(t)  \,.
\end{eqnarray}
This observable has a  different interpretation depending on whether the dynamics are under- or overdamped, both of which we consider below. In the underdamped case, the current is simply a function of the state with $A(\bf{x})=v$ and $B=0$, while in the overdamped case it depends on the stochastic increment, $A(\bf{x})=0$, $B(\bf{x})=1$.

\begin{figure*}[t]
	\includegraphics[width=1\linewidth]{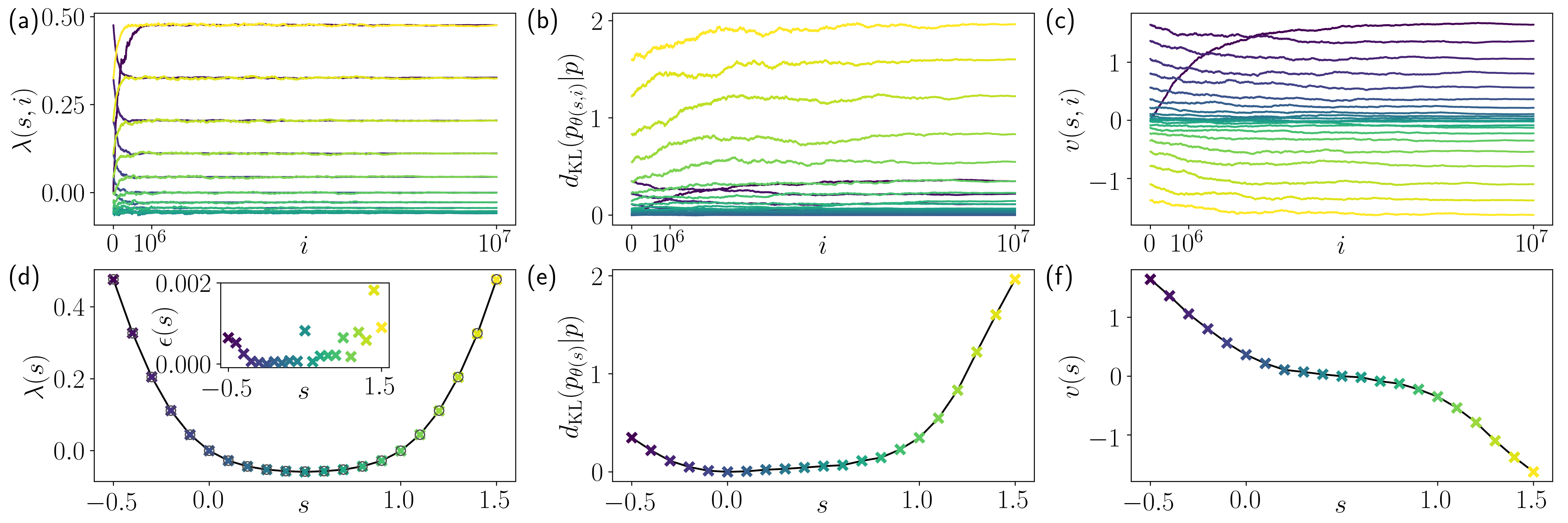}
	\caption{\label{fig:overdamped-learning} \textbf{Overdamped current fluctuations:} (a) Learning curves showing running estimates of the SCGF, (b) time-averaged KL divergence to the original dynamics $d_\mathrm{KL}(p_{\theta(s,i)}|p)$ during training for bias $s$ at step $i$, and (c) the time-averaged velocity. The color of each curve indicates the value of of the bias $s$, corresponding with the colors of the data points in the lower plots. Estimates of (d) the SCGF, (e) time-averages KL divergence with the original dynamics and (f) time-averaged velocity, for the final dynamics found at each value of the bias $s$ indicated on the $x$ axis. The inset of (d) shows the absolute error with numerical diagonalization results, represented by grey circles in (d).}
\end{figure*}
The corresponding scaled CGF we aim to compute is
\begin{equation}
\lambda(s) = \lim_{T\rightarrow\infty}\frac{1}{T} \ln  \left \langle e^{-s J[\bf{X}_{0,T}]} \right \rangle_{p} .
\end{equation} 
The first derivative of $\lambda(s)$ 
\begin{equation}
v(s) = -\frac{d \lambda(s)}{d s}
\end{equation}
reports on the average velocity in the tilted ensemble and is a useful indicator of the tails of the reference distribution. The scaled CGF exhibits a Lebowitz-Spohn symmetry\cite{lebowitz1999gallavotti} such that 
\begin{equation}
\lambda(s) =\lambda(-f-s) 
\end{equation}
where $f$ is the affinity for the current. The scaled CGF can be computed by the numerical solution of a generalized eigenvalue problem,\cite{Das2019} which we use for this low dimensional system to compare the accuracy of our results. 

Despite its simplicity this system has been shown to present non-trivial non-equilibrium phenomena due to the competition between ballistic and diffusive motion \cite{TsobgniNyawo2016,Ma2017,Fischer2018}.
Here, the overdamped regime acts as a simple benchmark which can be easily solved by diagonalizating a projection of the Fokker-Planck equation \cite{Fischer2018}.
The underdamped regime is a much more difficult problem to solve, due to a higher dimensional state space and long relaxation time.
Indeed, despite access to the SCGF via diagonalization \cite{Fischer2018}, accurate results for the force in the underdamped case have been elusive.
However, the actor-critic approach can solve this problem easily.

\begin{figure*}[t]
	\includegraphics[width=1\linewidth]{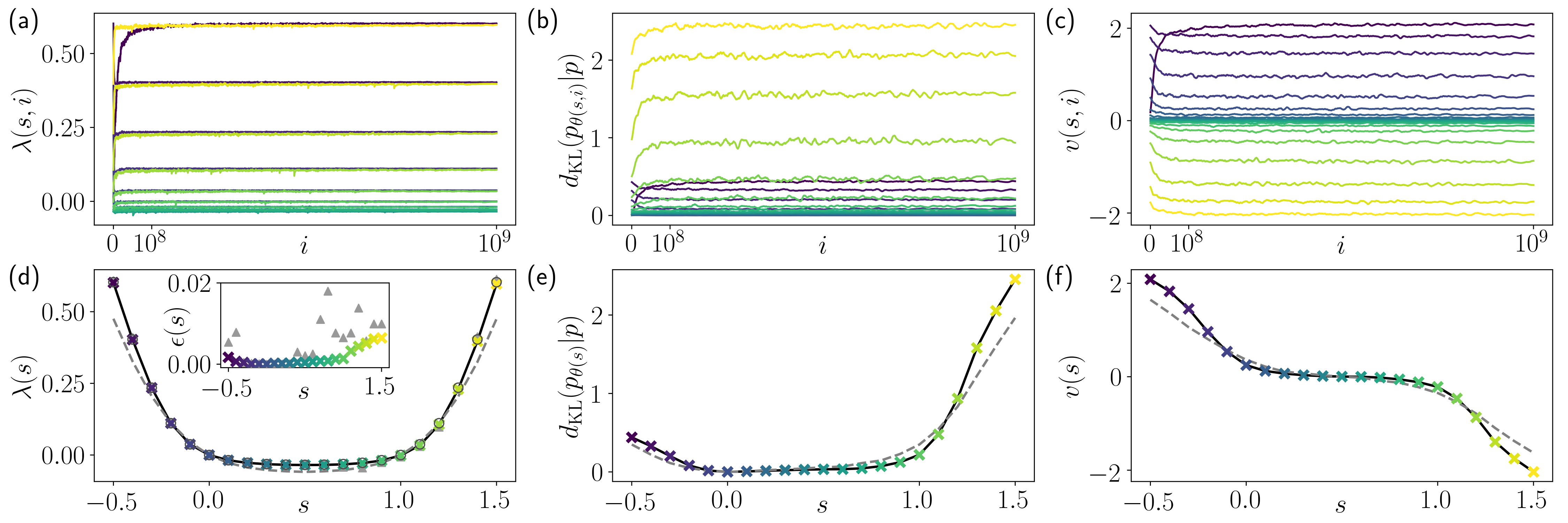}
	\caption{\label{fig:underdamped-learning} \textbf{Underdamped current fluctuations:} (a) Learning curves showing running estimates of the SCGF, (b) time-averaged KL divergence to the original dynamics $d_\mathrm{KL}(p_{\theta(s,i)}|p)$ during training for bias $s$ at step $i$, and (c) the time-averaged velocity, calculated as the dynamics is trained. The color of each curve indicates the value of of the bias $s$, corresponding with the colors of the data points in the lower plots. Estimates of (d) the SCGF, (e) time-averages KL divergence with the original dynamics and (f) time-averaged velocity of the final dynamics for each value of the bias $s$ indicated on the $x$ axis. The inset of (d) shows the absolute error with numerical diagonalization results, represented by grey circles in (d). Results with estimated corrections using the algorithm in Ref.\onlinecite{Das2019} are shown as triangles in (d) and its inset. Dashed curves in (d-f) show the results for the overdamped case for comparison.}
\end{figure*}
\subsection{Current fluctuations of an overdamped particle}\label{sec:overdamped}
 In the overdamped case, the evolution equation for the particle on a ring is given by
\begin{eqnarray}
    dx &=& F(x)dt + \sqrt{2}dW
\end{eqnarray}
which is a dimensionless one-dimensional SDE. We integrate this equation with a timestep of 0.001. Since the position is periodic, an ideal representation of both the force and value function is given by a Fourier series
\begin{eqnarray}
    F_{\theta}(x)=F(x) + a^\theta+\sum_{i=1}^M b^\theta_i\sin(ix)+c^\theta_i\cos(ix),
\end{eqnarray}
and 
\begin{eqnarray}
    V_{\psi}(x)=a^\psi+\sum_{i=1}^M b^\psi_i\sin(ix)+c^\psi_i\cos(ix),
\end{eqnarray}
with coefficients $a^\theta$,$a^\psi$, $\{b_i^\theta,c_i^\theta\}_{i=1}^M$ and $\{b^\psi_i,c^\psi_i\}_{i=1}^M$ truncated to dimension $M$.

The results of the differential AC algorithm are shown in Fig.~\ref{fig:overdamped-learning}.
We have truncated the basis with $M=5$ and used learning rates of $\alpha_\theta=0.1$ and $\alpha_\psi=0.01$. 
We annealed across the range  $s$ considered, first learning the dynamics at $s=-0.5$, before sweeping across to $s=1.5$ in steps of $\Delta s=0.1$.
The reward learning rate began at $\alpha_R=10^{-5}$ and decreased linearly to $\alpha_R=10^{-6}$ throughout training at each value of $s$, to enable rapid convergence to an accurate result.

We detail estimates of three quantities calculated during the learning process. In Fig.~\ref{fig:overdamped-learning}\textcolor{blue}{(a)} we show the estimate of $\lambda(s)$, the quantity the algorithm is attempting to maximize. In Fig.~\ref{fig:overdamped-learning}\textcolor{blue}{(b)} we show an estimate of the time-averaged KL divergence. In Fig.~\ref{fig:overdamped-learning}\textcolor{blue}{(c)} we show an estimate of the time-averaged velocity.
These estimates are running averages calculated using the samples taken from each transition, with learning rates of $0.1\alpha_R$.
Learning curves are plotted for training at each individual bias $s$ during the annealing process.
For small changes of $s$, we see that convergence to an accurate estimate of the scale CGF is achieved in approximately $10^6$ training steps, each utilizing data from a single transition. This results in a speed of up to two orders of magnitude over the MCR algorithm \cite{Das2019}.

In Figs.~\ref{fig:overdamped-learning}\textcolor{blue}{(d-f)} we plot the end points of each of these learning curves for the three observables plotted in Figs.~\ref{fig:overdamped-learning}\textcolor{blue}{(a-c)}.
In Fig.~\ref{fig:overdamped-learning}\textcolor{blue}{(d)} we see the expected Lebowitz-Spohn symmetry with reflection about $s=1/2$ for the scaled CGF. The inset shows the absolute error compared to the diagonalization of the Fokker-Planck equation, $\epsilon(s)$, which illustrates quantitative accuracy across the $s$ values considered. The maximal error is on the order of 1$\%$. 
Likewise, we see the expected anti-symmetry in the time-averaged KL divergence and velocity in Figs.~\ref{fig:overdamped-learning}\textcolor{blue}{(e) and (f)}. Both of these are also quantitatively accurate. 
This antisymmetry  implies that the optimal force differs from the reference force more for $s>1$ than $s<0$. This demonstrates that the regular production of trajectories with significant negative time-integrated velocities requires a substantial change in the systems dynamics, in contrast to those with a significant positive velocity. Nevertheless the learning algorithm employed here is capable of parametrizing the modified force sufficiently well to work across these regimes.

\subsection{Current fluctuations of an underdamped particle}\label{sec:underdamped}

In the underdamped case, the position and velocity evolve according to two coupled SDEs given by
\begin{eqnarray}
    dx &=& v dt,\\
    dv &=& F(x)dt-vdt + \sqrt{2}dW \nonumber
\end{eqnarray}
where the noise acts only on the velocity, $v$, and the friction, inverse temperature, and mass are taken as unity. As before we discretize our equations with a timestep of 0.001. 
For the underdamped case, the modified force and value function depends on both the position and velocity of the particle.
The approximation need only provide a single output for a force applied to the velocity, as the optimal dynamics can not change the evolution of the position since the position is not directly influenced by noise. To do accomplish this, a simple approach we have taken is to discretize the force and value function approximation along the velocity dimension.
More precisely, we can adapt the Fourier series from the overdamped case,
\begin{equation}
    F_\theta(x,v)=a^\theta(v)+\sum_{i=1}^{M_1}b^\theta_i(v)\sin(ix)+c^\theta_i(v)\cos(ix),
\end{equation}
with velocity dependent coefficients given by
\begin{eqnarray}
    a^\theta(v)&=&a_0 I_0(v)+a_{M_2+1}I_{M_2+1}(v)+\sum_{j=1}^{M_2}a_j I_{j,j+1}(v)
\nonumber\\
%    &&,
\end{eqnarray}
where 
\begin{equation}
    I_{j,j+1}(v) = \begin{cases} 1 \quad v_0+j\Delta v < v < v_0+(j+1)\Delta v \\ 0 \quad \mathrm{else} \end{cases}
\end{equation}
and the boundary cases $I_0(v)$ and $I_{M_2+1}(v)$ return 1 for $v$ less than $v_0$ or greater than $v_0+(M_2+1)\Delta v$, respectively. We employ analogous equations for $b^{\theta}_i(v)$ and $c^{\theta}_i(v)$. To achieve accurate results, we find a spacing of $\Delta v=0.02$ is sufficient, with $v_0=-8$, ${M_2}=700$ providing a broad enough range to encompass all relevant velocities at the biases considered. We use a Fourier basis with $M_1=5$. As before, we use the same functional for the value function as for this modified force. 

Figure \ref{fig:underdamped-learning} shows estimates of same three quantities as the overdamped case throughout the same annealed learning process.
Here we increased the value learning rate to $\alpha_\psi=0.1$, retain a dynamics learning rate of $\alpha_\theta=0.1$, and keep the scaled CGF learning rate fixed to $\alpha_R=10^{-6}$ throughout training.
Curves in Figs.~\ref{fig:underdamped-learning}\textcolor{blue}{(b,c)} are produced from data calculated using the same learning rate as the scaled CGF, before using a windowed average over $100$ steps to smooth the curve.
We generally see fast convergence to an accurate result in approximately $10^8$ transitions worth of updates. The large learning time compared to the overdamped results reflect the significantly finer basis employed for the underdamped model. 

The ends of these curves are plotted below in Figs.~\ref{fig:underdamped-learning}\textcolor{blue}{(d-f)}.
In the inset of Figs.~\ref{fig:underdamped-learning}\textcolor{blue}{(d)} we see that we find accurate results compared to the numerically exact answers across the range of $s$ considered. 
We see analogous results to the overdamped case, reproduced in dashed lines in Figs.~\ref{fig:underdamped-learning} \textcolor{blue}{(e,f)}, the underdamped system obeys the expected Lebowitz-Spohn symmetry. Compared to the overdamped system, the features of the KL divergence and average velocity in underdamped system are sharper.

\begin{figure}[t]
	\includegraphics[width=1\linewidth]{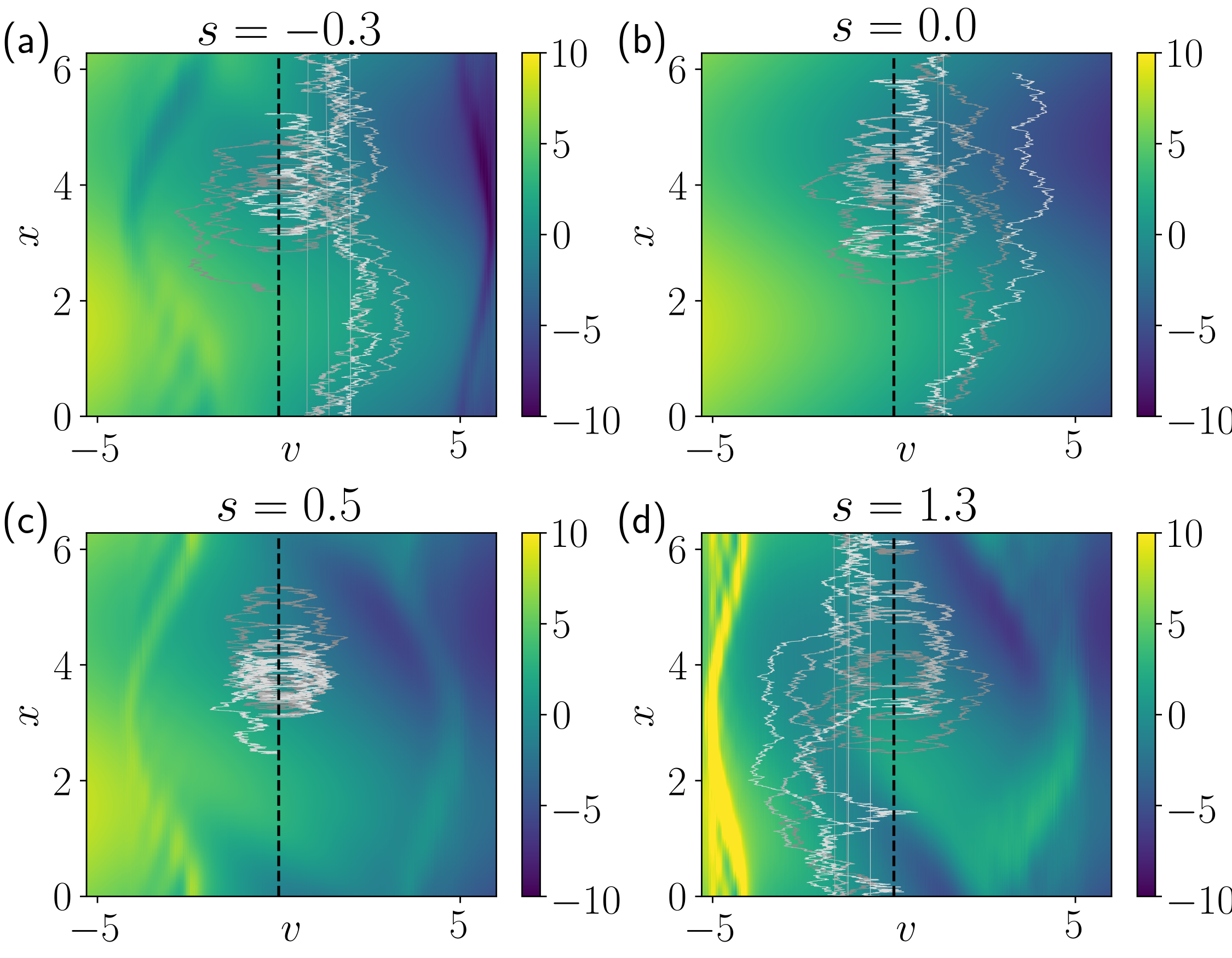}
	\caption{\label{fig:underdamped-forces} \textbf{Modified forces and their dynamics:} The final forces learnt during the optimization process for bias $s=-0.3$(a), $0.0$(b), $0.5$(c) and $1.3$(d) with $3$ sample trajectories of length $T=10$ for each force.}
\end{figure}

There are three distinct behaviors for the system as a function of $s$. For large negative $s$, the velocity increases significantly. For very large positive $s$ the velocity decreases analogously. For small and intermediate positive $s$, there is a broad plateau where the velocity is close to zero.  These distinct regions are clearly demonstrated in Fig \ref{fig:underdamped-forces} where we plot the final optimized forces for a set of $s$, along with sample trajectories generated by these forces.
We see different behavior for biases of $s<0$, $0<s<1$ and $1<s$. For $s<0$ the trajectories  regularly loop round the ring in the positive direction. For $0<s<1$ the trajectories generally do not transition round the ring and instead remain in a small region of space. For $s>1$ the trajectories loop around the ring in the negative direction.

\begin{figure}
	\includegraphics[width=1\linewidth]{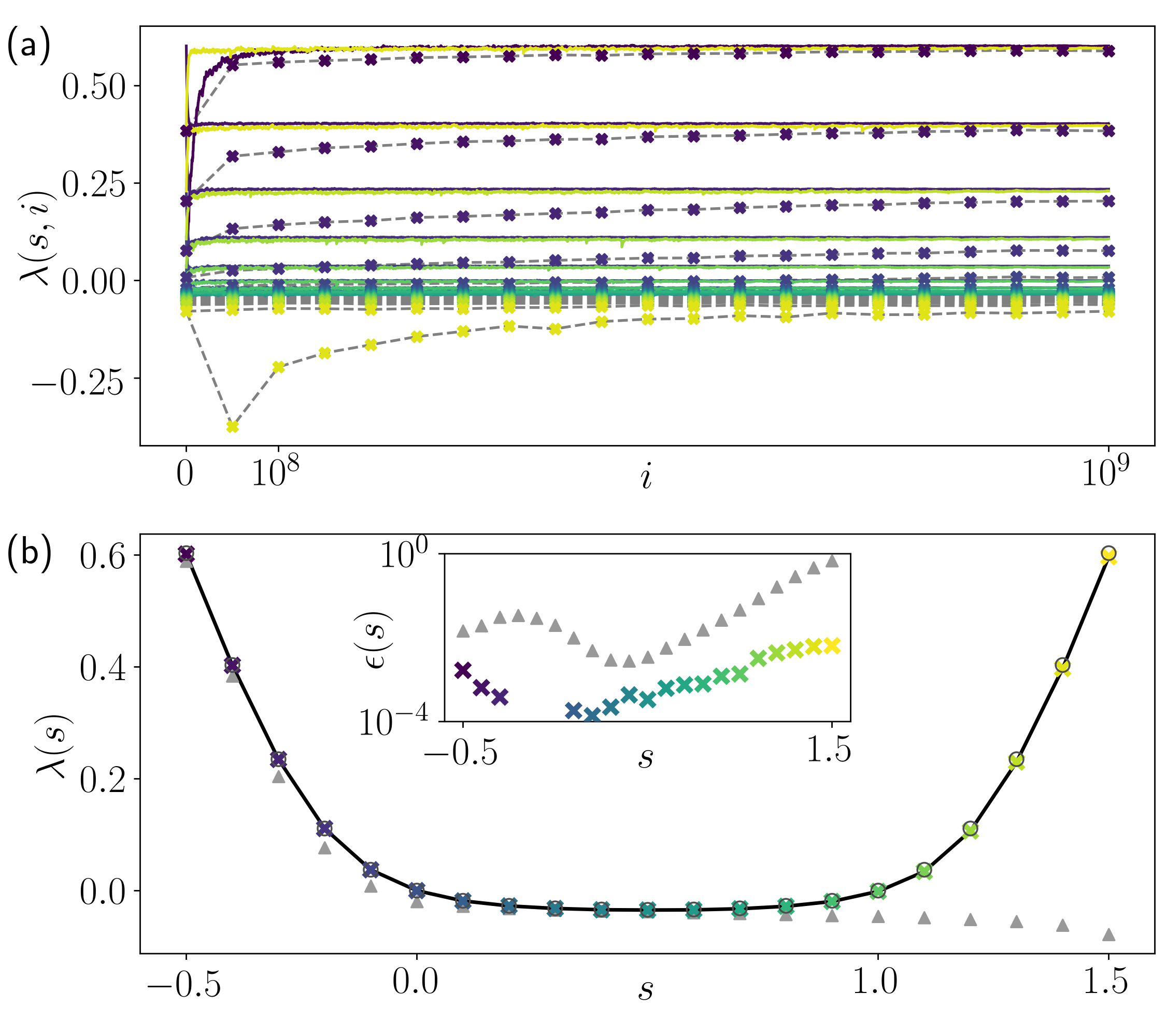}
	\caption{\label{fig:underdamped-comparison} \textbf{Comparison between AC and MCR algorithms:} (a) learning curves plotted verses the amount of data used during training, for the AC algorithm (solid, colored lines) and the MCR algorithm  (colored crosses and dashed gray lines). Curves and crosses are color coded by the value of the bias $s$ being trained for. (b) Final results for the AC algorithm (colored crosses) and the MCR algorithm (gray triangles), with absolute errors to the value from numerical diagonalization shown in the inset.}
\end{figure}

For comparison, we have optimized the same functional form using the MCR algorithm, as analogous to Ref.~\onlinecite{Das2019}. The AC algorithm provides more accurate results than MCR, when optimized using the same amount of statistics \cite{Das2019}. The MCR results are produced by annealing across from $s=1.5$ down to $s=-0.5$ in steps of $0.1$. Training for each value of $s$ involves $20$ updates constructed using $50$ trajectories with $10^6$ time steps each, for a total of $10^9$ transitions worth of data.
After optimizing the hyperparameters, we see in Fig.~\ref{fig:underdamped-comparison} the convergence in the MCR algorithm is still much slower than the AC algorithm. As a consequence, the best results we can achieve using the same amount of transitions fail to converge to the correct values of the scaled CGF for biases close to $s\gtrsim1$.
This demonstrates one key advantage of utilizing value functions.  Due to the reduction in variance of gradient estimates using a small amount of data, we can perform many more updates using the same amount of transitions, improving convergence.


\section{Conclusions}\label{sec:conclusion}
In this paper we have demonstrated how regularized reinforcement learning algorithms can be used to optimize a diffusive dynamics to effectively sample rare trajectories.
A key ingredient of our approach is a value function that estimates how relevant each state is to the rare dynamics, a function learnt while simultaneously guiding optimization of the dynamics, allowing for reduced data generation and more detailed function approximations. Across a range of systems and observables, we found that the lower variance estimate of the gradient employing value functions enabled accurate and efficient characterization of rare dynamical fluctuations. In finite time problems, the AC algorithm in particular was able to solve particularly challenging problems associated with multiple reactive channels and long lived intermediates. In the long time limit, the AC algorithm reproduces exact results for the cumulant generating function by directly optimizing to an accurate representation of the Doob dynamics, removing the need to calculate additional corrections or do additional importance sampling.

While we have focused here on the simulation of rare event dynamics and the direct evaluation of their likelihoods, the methods of finding optimized forces developed here can be straightforwardly combined with trajectory importance sampling methods such as transition path sampling\cite{Bolhuis2002} or cloning\cite{giardina2006direct} to correct for inaccuracies associated with an incomplete basis. Indeed, previous work has demonstrated that auxiliary dynamics can significantly improve the statistical efficiency of trajectory sampling methods.\cite{Das2019,Ray2018a,nemoto2017finite,bartolucci2018transition} Further, Monte Carlo approaches can be used to generate data to train the optimal dynamics in a feedback routine as previously demonstrated.\cite{Nemoto2016,Oakes2020} This could emphasize the parts of the state space relevant to the rare events earlier than by simply generating data with the current dynamics, thus speeding up optimization. Application to more complex models, such as many-body systems, will be an important development of this line of research. Accurate approximation of the force in many-body problems may require the use of more sophisticated function approximations, such as neural networks, however, a difficult balance will need to be struck between the representative power of the approximation and the computational cost to calculate it. More powerful function approximations will also necessitate the use of more sophisticated algorithms, as training such approximations can become unstable when using correlated data, as we do here.

\acknowledgments
AD and DTL were supported by NSF Grant CHE1954580. DCR and JPG were supported by University of Nottingham grant no.\ FiF1/3 and EPSRC Grant no.\ EP/R04421X/1. JPG is grateful to All Souls College, Oxford, for support through a Visiting Fellowship during part of this work. DCR is grateful for access to the University of Nottingham Augusta HPC service. 

\section*{Data Availability} The data that support the findings of this study are openly available in Zenodo at \href {\doibase 10.5281/zenodo.4772483}{https://doi.org/10.5281/zenodo.4772483}.\cite{dcrose_adas_2021_4772483}

\appendix

\section{Discrete timestep implementations of finite time algorithms}\label{sec:discrete-timestep}

We now describe how the time-continuous equations of the reinforcement learning algorithm are efficiently implemented in simulations with a fixed discrete timestep $\Dt$, though variable timesteps may be easily used. We use an Euler propagator to integrate the SDE in Equation (\ref{eq:modifiedsde}) as
\begin{equation}\label{eq:discrete-evolution}
    \bf{x}_{t+\Dt} = \bf{x}_{t} + \Dt \bf{F}_{\theta}(\bf{x}_{t},t) + \G\Delta\bf{W}_{t}
\end{equation}
where $\Delta\bf{W}$ is a Gaussian random variable with mean $0$ and variance $\Dt$. The trajectory probability from Eq.~(\ref{eq:parametrized-trajectory-prob}) is now given by products of stepwise probabilities
\begin{align}
	&p_\theta\left[\bf{X}_{t,t+\Dt}|\bf{x}_t\right]\nonumber\\
	&=\frac{\exp\left \{-\frac{1}{2\Dt}\left |\G^{-1} \left(\bf{x}_{t+\Dt}-\bf{x}_{t} - \Dt\bf{F}_{\theta}(\bf{x}_{t},t)\right)\right |^2\right\} }{2\pi\Dt\;\mathrm{det}(\G)}\, 
\end{align}
Next we discretize the gradient of the logarithm of trajectory probabilities using the Ito convention. We propagate the Malliavin weights from Eq.~(\ref{eq:malliavinweightprop}) as
\begin{align}
    y_{\theta}(t+\Dt)=y_{\theta}(t)+&\left[\G^{-1}\left( \bf{x}_{t+\Dt}-\bf{x}_{t}- \Dt\bf{F}_\theta(\bf{x}_{t},t)\right)\right] \nonumber\\
    \cdot&\left[ \G^{-1}\gradt \bf{F}_\theta(t)\right]
\end{align}
We also write the full return (\ref{eq:returnfull}) through a sum of stepwise rewards as
\begin{align}
	R\left[\bf{x}_{t^-,t+\tau}\right]=\sum_{j:j\Dt<\tau}r\left( \bf{x}_{j+1},\bf{x}_{j},t+j\Dt\right)
\end{align}
where the timestep index $j$ starts from -1 in this sum, with the notation $t^{-}$ accounting for the timestep before the current one, and the subscript $j$ refers to the time $t+j\Dt$. The reward at each step is defined as
\begin{align}
	&r\left(\bf{x}_{j+1},\bf{x}_{j},t+j\Dt\right)\nonumber\\
	&=-s\left(A_j\Dt+\bf{B}_j \cdot(\bf{x}_{j+1}-\bf{x}_j)+A(\bf{x}_{j+1})\delta_{jn}\right)\nonumber\\
	&+\frac{\left[\G^{-1}(\bf{x}_{j+1} - \bf{x}_{j} - \Dt \bf{F}_{\theta}(\bf{x}_{j},t_{j}))\right]^2}{2}\nonumber\\
	&-\frac{\left[\G^{-1}(\bf{x}_{j+1} - \bf{x}_{j} - \Dt \bf{F}(\bf{x}_{j},t_{j}))\right]^2}{2},
\end{align}
using the definition of the observable from Eq.~(\ref{eq:time-integrated-observable}) and accounting for an additional singular reward at the end of the trajectory after the last timestep $n$. Here the first three terms come from the observable and the last two terms represent the KL divergence between the original and optimized dynamics.

Now we combine the rewards, Malliavin weights and value functions in multiple ways to produce the gradients in the different algorithms. The pseudocodes of efficient implementations of these are presented below.

%\begin{align}
%	&\gradt\DKL\approx\nonumber\\
%	&-\sqrt{\Dt}\left\langle\sum_{t:t=i\Dt}\delta_{\psi_i}\left[x_{t^-,t+\tau},t\right]w_i^TG_i^{-1}\gradt F^\theta_i\right\rangle_{p_\theta},\nonumber\\
%	&\left.\gradp L(\psi,\psi_i)\right|_{\psi=\psi_i}\nonumber\\
%	&=-\Dt\left\langle\sum_{t:t=i\Dt}\delta_{\psi_i}\left[x_{t^-,t+\tau},t^-\right]
%	\left.\gradp V_\psi\left[x(t),t\right]\right|_{\psi=\psi_i}\right\rangle_{p_\theta}.
%\end{align}

\subsection{Monte-Carlo returns}
The gradient in the Monte Carlo returns algorithm can be rewritten from Equation (\ref{eq:mcr}) as
\begin{align}
    \chi_{\mathrm{MCR}}(\theta,T)&=-\left\langle \int_{0}^{T}dt\;R\left[ \bf{X}_{t^{-},T}\right] \dot{y}_{\theta}(t)\right\rangle _{p_{\theta}}\nonumber\\
    &=-\left\langle \int_{0}^{T}dt\; \dot{y}_{\theta}(t)\int_{t^{-}}^{T}dt^{'}\;\dot{R}(t^{'})\right\rangle _{p_{\theta}}\nonumber\\
    &=-\left\langle \int_{0}^{T}dt\;\dot{R}(t) \int_{0}^{t^{+}}dt^{'}\;\dot{y}_{\theta}(t^{'})\right\rangle _{p_{\theta}}\nonumber\\
    &=-\left\langle \int_{0}^{T}dt\;\dot{R}(t)y_{\theta}(t^{+})\right\rangle _{p_{\theta}}
\end{align}
where the return has been written as a time integral of its differential changes, and $t^{+}$ is shorthand for $t+\epsilon$ for some small positive $\epsilon$. This has converted the double time integral into a single time integral, which is then evaluated on-the-fly while propagating the trajectory. An implementation of this algorithm with a fixed timestep $\Dt$ is described in the pseudocode in Alg. \ref{mcr-algo}.

\begin{algorithm}[H]
	\caption{Finite time MCR}\label{mcr-algo}
	\begin{algorithmic}[1]
		\State \textbf{inputs} dynamical approximation $\bf{F}_{\theta}(\bf{x},t)$
		\State \textbf{parameters} learning rate $\alpha^\theta$; total optimization steps $I$; trajectory length $T$ consisting of $J$ timesteps of duration $\Dt$ each; number of trajectories $N$
		\State \textbf{initialize} choose initial weights $\theta$, define iteration variables $i$ and $j$, force gradient $\delta_{P}$, stepwise rewards $r$ representing the increments in return
		\State $i\gets0$
		\Repeat
		\State Using chosen method to generate trajectories $\bf{X}_{0,T}$ with configurations, times, noises, Malliavin weights and rewards denoted by $\bf{x}_{j},t_{j},\Delta\bf{W}_{j},y_{\theta}(t_{j})$ and $r(\bf{x}_{j+1},\bf{x}_{j},t_{j})=r_{j}$ respectively
		\State $j\gets0$
		\State $\delta_P\gets0$		\State $y_{\theta}(t_{0})\gets0$
		\Repeat
		\State $y_{\theta}(t_{j+1})\gets y_{\theta }(t_{j})+\Delta\bf{W}_{j}\cdot[\G^{-1}\nabla_{\theta}\bf{F}_{\theta}(\bf{x}_{j},t_{j})]$
		\State $\delta_P\gets\delta_P+r_{j}y_{\theta}(t_{j+1})$
		\State $j\gets j+1$
		\Until{$j=J$}
		\State average $\delta_{P}$ over $N$ trajectories to get $\overline{\delta}_{P}$
		\State $\theta\gets\theta+\alpha^\theta\overline{\delta}_P$
		\State $i\gets i+1$
		\Until{$i=I$}
	\end{algorithmic}
\end{algorithm}

\subsection{Monte-Carlo returns with a value baseline}
We use a similar technique to rewrite the double time integral for the gradient in the Monte Carlo value baseline algorithm, Equation (\ref{eq:mcvb-grad}), using a single time integral as
\begin{align}
    &\chi_{\mathrm{MCVB}}(\theta,T)\nonumber\\
    &=-\left\langle \int_{0}^{T}dt\;\left\{ R\left[ \bf{X}_{t^{-},T}\right] -V_{\psi}(\bf{x}_{t},t)\right\} \dot{y}_{\theta}(t)\right\rangle _{p_{\theta},\psi=\psi_{i}}\nonumber\\
    &=-\left\langle \int_{0}^{T}dt\;\left\{ \dot{R}(t)y_{\theta}(t^{+})-V_{\psi}(\bf{x}_{t},t)\dot{y}_{\theta}(t)\right\} \right\rangle _{p_{\theta},\psi=\psi_{i}}.
\end{align}
We rewrite the gradient of the value error in Eq. (\ref{eq:vgrad}) similarly as
\begin{align}
    \gradp L(\psi,\psi_i)&\biggr|_{\psi=\psi_i}\nonumber\\
	=-\Biggl\langle \int_{0}^{T}dt\;&\biggl\{ \dot{R}(t)\left( \int_{0}^{t^{+}}dt^{'}\nabla_{\psi}V_{\psi}(t^{'})\right) \nonumber\\
	&-V_{\psi}(t)\nabla_{\psi}V_{\psi}(t)\biggr\} \Biggr\rangle_{p_\theta, \psi=\psi_{i}}\nonumber\\
	=-\Biggl\langle \int_{0}^{T}dt\;&\biggl\{ \dot{R}(t)z_{\psi}(t^{+}) -V_{\psi}(t)\dot{z}_{\psi}(t)\biggr\} \Biggr\rangle_{p_\theta, \psi=\psi_{i}} ,
\end{align}
where the arguments of the value function $V_{\psi}(\bf{x}_{t},t)$ have been suppressed as $V_{\psi}(t)$ and the integral of the gradient of the value function upto and including current time has been denoted as $z_{\psi}(t^{+})$. We explicitly set the $V(\bf{x}_{t},t)$ to 0 for any $t\geq T$, i.e., after the last timestep, in these expressions. The single time integral is then evaluated on-the-fly as the trajectory is propagated. If the force and the value function approximations use the same set of basis functions as we do with a fixed grid of Gaussians, the MCVB algorithm incurs no additional computational cost over the MCR algorithm. An implementation of this algorithm with a fixed timestep $\Dt$ is described in the pseudocode in Alg. \ref{mcvb-algo}.

\begin{figure}
\begin{algorithm}[H]
	\caption{Finite time MCVB}\label{mcvb-algo}
	\begin{algorithmic}[1]
		\State \textbf{inputs} dynamical approximation $\bf{F}_{\theta}(\bf{x},t)$, value approximation $V_{\psi}(\bf{x},t)$
		\State \textbf{parameters} learning rates $\alpha^\theta$, $\alpha^{\psi}$; total optimization steps $I$; trajectory length $T$ consisting of $J$ timesteps of duration $\Dt$ each; number of trajectories $N$
		\State \textbf{initialize} choose initial weights $\theta$ and $\psi$, define iteration variables $i$ and $j$, force and value function gradients $\delta_{P}$, $\delta_{V}$, stepwise rewards $r$ representing the increments in return
		\State $i\gets0$
		\Repeat
		\State Using chosen method to generate trajectories $\bf{X}_{0,T}$ with configurations, times, noises, Malliavin weights, integral of value function gradients, and rewards denoted by $\bf{x}_{j},t_{j},\Delta\bf{W}_{j},y_{\theta}(t_{j}),z_{\psi}(t_{j})$ and $r(\bf{x}_{j+1},\bf{x}_{j},t_{j})=r_{j}$ respectively
		\State $j\gets0$
		\State $\delta_P\gets0$
		\State $\delta_V\gets0$
		\State $y_{\theta}(t_{0})\gets0$
		\State $z_{\psi}(t_{0}\gets0)$
		\Repeat
		\State $\dot{y}_{\theta}(t_{j})\gets\Delta\bf{W}_{j}\cdot[\G^{-1}\nabla_{\theta}\bf{F}_{\theta}(\bf{x}_{j},t_{j})]/\Dt$
		\State $y_{\theta}(t_{j+1})\gets y_{\theta}(t_{j})+\Dt\dot{y}_{\theta}(t_{j})$
		\State $\dot{z}_{\psi}(t_{j})\gets\nabla_{\psi}V_{\psi}(\bf{x}_{j},t_{j})$
		\State $z_{\psi}(t_{j+1})\gets z_{\psi}(t_{j})+\Dt\dot{z}_{\psi}(t_{j})$
		\State $\delta_P\gets\delta_P+r_{j}y_{\theta}(t_{j+1})-V_{\psi}(\bf{x}_{j},t_{j})\dot{y}_{\theta}(t_{j}))$
		\State		$\delta_{V}\gets\delta_{V}+r_{j}z_{\psi}(t_{j+1})-V_{\psi}(\bf{x}_{j},t_{j})\dot{z}_{\psi}(t_{j})$
		\State $j\gets j+1$
		\Until{$j=J$}
		\State average $\delta_{P}$,$\delta_{V}$ over $N$ trajectories to get $\overline{\delta}_{P}$, $\overline{\delta}_{V}$
		\State $\theta\gets\theta+\alpha^\theta\overline{\delta}_P$
		\State $\psi\gets\psi+\alpha^{\psi}\overline{\delta}_{V}$
		\State $i\gets i+1$
		\Until{$i=I$}
	\end{algorithmic}
\end{algorithm}
\end{figure}

\subsection{Actor-critic}
We rewrite the gradient in the Actor-critic algorithm from Equation (\ref{eq:ac-grad}) using a shift in time origin as
\begin{align}
    &\chi_{\mathrm{AC}}(\theta,T)\nonumber\\
    =&-\left\langle \int_{0}^{T}dt\;\delta^{'}\left[ \bf{X}_{t^{-},t+\tau},t\right] \dot{y}_{\theta}(t)\right\rangle _{p_{\theta},\psi=\psi_{i}}\nonumber\\
    =&-\left\langle \int_{\tau}^{T+\tau}dt\;\delta^{'}\left[ \bf{X}_{t^{-}-\tau,t},t-\tau\right] \dot{y}_{\theta}(t-\tau)\right\rangle _{p_{\theta},\psi=\psi_{i}}
\end{align}
where the change in return and the value function for $t\geq T$ is explicitly set to 0. We similarly write the gradient of the value error from Eq. (\ref{eq:vgrad}) as 
\begin{align}
    \gradp L(\psi,\psi_i)\biggr|_{\psi=\psi_i}=-\Biggl\langle& \int_{\tau}^{T+\tau}dt\;\delta^{'}\left[ \bf{X}_{t^{-}-\tau,t},t-\tau\right] \nonumber\\
	&~\nabla_{\psi}V_{\psi}(\bf{x}_{t-\tau},t-\tau)\Biggr\rangle _{p_{\theta},\psi=\psi_{i}}
\end{align}
These integrals are then evaluated on-the-fly along with trajectory propagation. Since the gradients involve correlations of the differential return $r$ with the differential Malliavin weight $\dot{y}_{\theta}$ and the value function gradient $\dot{z}_{\psi}=\nabla_{\psi}V_{\psi}$ from $\tau$ time in the past, this makes it necessary to store and use this history, along with the reward and the value function, for the past $\tau/\Dt$ timesteps. Aside from this additional memory requirement, given a delay time $\tau$ which is much smaller than the trajectory duration, the Actor-critic algorithm has similar computational cost comparable to the MCR and MCVB algorithms. This implementation of the algorithm is described in the pseudocode in Alg. \ref{ac-algo}.

\begin{algorithm}[H]
	\caption{Finite time AC}\label{ac-algo}
	\begin{algorithmic}[1]
		\State \textbf{inputs} dynamical approximation $\bf{F}_{\theta}(\bf{x},t)$, value approximation $V_{\psi}(\bf{x},t)$
		\State \textbf{parameters} learning rates $\alpha^\theta$, $\alpha^{\psi}$; total optimization steps $I$; trajectory length $T$ consisting of $J$ timesteps of duration $\Dt$ each; temporal delay $M=\tau/\Dt$; number of trajectories $N$
		\State \textbf{initialize} choose initial weights $\theta$ and $\psi$, define iteration variables $i$ and $j$, force and value function gradients $\delta_{P}$, $\delta_{V}$, stepwise rewards $r$ representing the increments in return
		\State $i\gets0$
		\Repeat
		\State Using chosen method to generate trajectories $\bf{X}_{0,T}$ with configurations, times, noises, changes in Malliavin weights, value function gradients, temporal difference, rewards and cumulative rewards denoted by $\bf{x}_{j},t_{j},\Delta\bf{W}_{j},\Delta{y}_{\theta}(t_{j}),\dot{z}_{\psi}(t_{j}),\delta^{'}_{j}$, $r(\bf{x}_{j+1},\bf{x}_{j},t_{j})=r_{j}$ and $R\left[ \bf{X}_{t_{j}-\tau,t_{j}}\right] =R_{j-M,j}$ respectively, and $r_{j}=V(\bf{x},t_{j})=0$ whenever $j<0$ or $j\geq J$
		\State $j\gets0$
		\State $\delta_P\gets0$
		\State $\delta_V\gets0$
		\State $R_{-M,0}\gets0$
		\Repeat
		\State $R_{j-M,j}\gets R_{j-M-1,j-1}+r_{j}-r_{j-M}$
		\If{$j<J$}
		    \State $\Delta y_{\theta}(t_{j})\gets\Delta\bf{W}_{j}\cdot[\G^{-1}\nabla_{\theta}\bf{F}_{\theta}(\bf{x}_{j},t_{j})]$
    		\State $\dot{z}_{\psi}(t_{j})\gets\nabla_{\psi}V_{\psi}(\bf{x}_{j},t_{j})$
    	\EndIf
    	\If{$j\geq M$}
    	    \State $\delta^{'}_{j}\gets V(\bf{x}_{j},t_{j})+R_{j-M,j}-V(\bf{x}_{j-M},t_{j-M})$
    	    \State $\delta_P\gets\delta_P+\delta^{'}_{j}\Delta y_{\theta}(t_{j-M})$
		    \State	$\delta_{V}\gets\delta_{V}+\delta^{'}_{j}\dot{z}_{\psi}(t_{j-M})$
		\EndIf
		\State $j\gets j+1$
		\Until{$j=J+M$}
		\State average $\delta_{P}$,$\delta_{V}$ over $N$ trajectories to get $\overline{\delta}_{P}$, $\overline{\delta}_{V}$
		\State $\theta\gets\theta+\alpha^\theta\overline{\delta}_P$
		\State $\psi\gets\psi+\alpha^{\psi}\overline{\delta}_{V}$
		\State $i\gets i+1$
		\Until{$i=I$}
	\end{algorithmic}
\end{algorithm}

\section{Comparing errors in gradient estimates} 
\label{sec:error}

\begin{figure}[t]
	\includegraphics[width=1\linewidth]{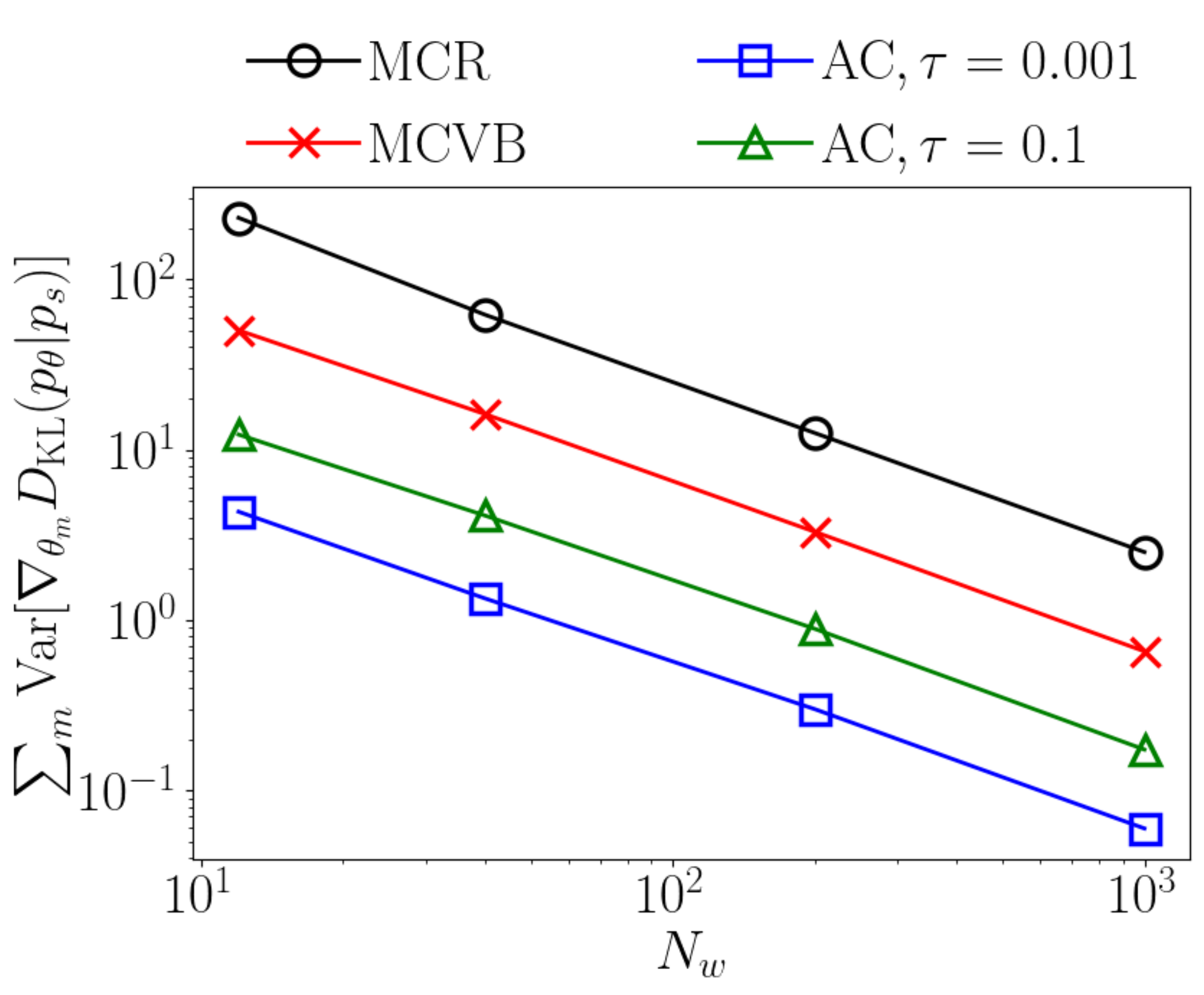}
	\caption {\textbf{Statistical convergence of gradient estimates} Total variance of the gradient summed over all components, using MCR(black), MCVB(red), AC with $\tau=0.001$(blue) and AC with $\tau=0.1$(green), as a function of the number of uncorrelated trajectories $N_{w}$ for averaging.}
	\label{fig:si1}
\end{figure}

In Figure \ref{fig:si1} we have directly compared the three algorithms for their ability to reduce the variance of the gradient estimates during optimization in the softened Brownian bridge problem. We have chosen
the force and value function coefficients $\theta$ and $\psi$ from the $i=100$ step of the MCVB optimization run in Fig. \ref{fig1}(b) in the Brownian bridge problem. This value function is thus not exact for the corresponding force but is representative of typical inaccuracies encountered during learning. Keeping these coefficients fixed, we have estimated the gradients of the KL divergence using the three algorithms, while varying the number of uncorrelated trajectories $N_{w}$ over which the estimates are averaged. Plotted in Fig. \ref{fig:si1} are the total variance in the gradient estimate summed over all components, $\sum_{m}\mathrm{Var}[\nabla_{\theta_{m}}D_{\mathrm{KL}}(p_{\theta}|p_{s})]$, from the different algorithms. The variances are computed from fluctuations over 10 uncorrelated sets of $N_{w}$ trajectories. The dependence on $N_{w}$ in log-log scale corresponds to a linear trend with a slope of $-1$ as expected from the variance of sample means of uncorrelated samples. We find that use of the MCVB and AC algorithms greatly reduces the variance compared to the MCR approach, equivalent to a 5 to 100 times increase in the amount of input trajectory data. We find that the smallest variance corresponds to the AC algorithm with the smallest possible $\tau$, set to the timestep $0.001$. However, this choice incurs a systematic error in the expectation of the gradient due to the inaccuracy in the value function, while neither MCVB nor AC with a large $\tau$ are susceptible to it. This is manifested in the scaled $L^{1}$ norm of the error in the expected gradient from the algorithms. The expectation is calculated over $10^{5}$ trajectories and the error in MCR is zero by definition. The $L^{1}$ norms of the errors, divided by that of the true gradient, are 0.22, 7.49 and 1.16 from MCVB, AC($\tau=0.001$) and AC($\tau=0.1$) respectively. This shows that the systematic error incurred by AC at small $\tau$ can be reduced by having a larger $\tau$, while still having significantly less variance than MCVB and MCR. The crossover between the systematic and statistical error in the AC algorithm depending on $\tau$ is also the reason starting the optimization with a small $\tau$ and later annealing with a large $\tau$ is an efficient strategy, given that the memory requirement scales linearly with $\tau$.
We note that the systematic error is formally zero by definition in the expectation of the MCVB gradient estimate as well: the small non-zero value stems from a finite number of samples being used to estimate the expectation.

\section{Alternative CGF estimates} 
\label{sec:altcgf}

 \subsection{Numerically exact CGF}

We have compared the CGF from the reinforcement learning algorithms in Section \ref{sec:2dL} with numerically exact values obtained from explicitly calculating $\langle h_{\Gamma}\rangle_{p}$ in equation (\ref{eq:fprobab}) by solving the corresponding Fokker-Planck operator. The Fokker-Planck operator for the original dynamics in Eq. (\ref{Eq:2dL}) is given by
 \begin{equation}
     L=-\nabla.\bf{F}(\bf{x})+\nabla^{2}
 \end{equation}
 where $\bf{F}(\bf{x})=-\nabla U(x)$ is the underlying conservative force.
 
 We want to use this operator in order to find the probability $\langle h_{\Gamma}\rangle_{p}$ as
 \begin{equation}
     \langle h_{\Gamma}\rangle_{p}=\int_{\Gamma}d\bf{x}\;\rho(\bf{x},T)=\int_{\Gamma}d\bf{x}\;e^{LT}\delta(\bf{x}-\bf{x}_{0})
 \end{equation}
 We exponentiate the operator in its spectral eigenbasis. Since the forces in the original dynamics are conservative, diagonalizing $L$ becomes easier through a similarity transform into a Hermitian operator $\mathcal{L}$,\cite{risken1996fokker,Majumdar2015}
 \begin{align}
     \mathcal{L}&=e^{U(\bf{x})/2}Le^{- U(\bf{x})/2}\nonumber\\
     &=\nabla^{2}-\frac{1}{4}(\nabla U(\bf{x}))^{2}+\frac{1}{2}\nabla^{2}U(\bf{x}).
 \end{align}
 We diagonalize $\mathcal{L}$ to obtain eigenvalues $-\lambda_{n}$ and eigenfunctions $\phi_{n}(\bf{x})$,
 \begin{equation}
     \mathcal{L}\phi_{n}(\bf{x})=-\lambda_{n}\phi_{n}(\bf{x}).
 \end{equation}
 Since $\mathcal{L}$ is Hermitian, the eigenfunctions $\{\phi_{n}(\bf{x})\}$ are mutually orthonormal and can be used to introduce a resolution of identity
 \begin{equation}
     \delta(\bf{x}-\bf{x}_{0})=\sum_{n}\phi_{n}(\bf{x}_{0})\phi_{n}(\bf{x})
 \end{equation}
 The original operator $L$ related by the similarity transform has eigenvalues $-\lambda_{n}$ and eigenfunctions $e^{-U(\bf{x})/2}\phi_{n}(\bf{x})$. This spectral expansion of $L$ can be used to estimate the probability $\langle h_{\Gamma}\rangle_{p}$ as
 \begin{align}\label{eq:spectral}
     \langle h_{\Gamma}\rangle_{p}&=\int_{\Gamma}d\bf{x}\;e^{LT}\delta(\bf{x}-\bf{x}_{0})\nonumber\\
     &=e^{U(\bf{x}_{0})/2}\sum_{n}e^{-\lambda_{n}T}\int_{\Gamma}d\bf{x}\;e^{-U(\bf{x})/2}\phi_{n}(\bf{x})
 \end{align}
 The final time $T$ that we use in our barrier-crossing simulations is chosen such that $\tau_{\mathrm{rlx}}<T<\tau_{\mathrm{rxn}}$ where $\tau_{\mathrm{rlx}}$ and $\tau_{\mathrm{rxn}}$ are respectively the timescale of relaxation in the starting or the ending well, and the timescale of the barrier-crossing reaction, which is expected to be the slowest dynamical mode in the system. Hence when the set $\{\lambda_{n}\}$ is ordered, the factor $e^{-\lambda_{n}T}$ should be negligible for all but the few smallest values of $n$. The sum over $n$ in Equation (\ref{eq:spectral}) is thus expected to converge within a few terms. 
 
We diagonalize the operator $\mathcal{L}$ using a Discrete Variable Representation basis constructed from Hermite polynomials\cite{szalay1993discrete} in two dimensions, $\chi_{M,N}(\alpha x,\alpha y)$, where $\alpha=5$ is a scaling factor. We obtain identically converged estimates of $\langle h_{\Gamma}\rangle_{p}$ with basis sizes ranging from $50\times50$ to $100\times100$ using 10 terms in the spectral expansion. The CGF value is then calculated using $\langle h_{\Gamma}\rangle_{p}$ in Equations (\ref{eq:cgfform}) and (\ref{eq:fprobab}).

\subsection{CGF from Kramers escape rate}

In one-dimension, corresponding to a dynamics of 
\begin{equation}
    dq=-U^{'}(q)+\sqrt{2}dW,
\end{equation}
an approximate expression for the barrier-crossing probability in time $T$ is given by the Kramers escape rate in the overdamped limit,\cite{zwanzig2001nonequilibrium} as
\begin{equation}
    \langle h_{\Gamma}\rangle_{p}\approx \frac{T}{2\pi}(U^{''}(q_{A})|U^{''}(q^{\dagger})|)^{1/2}e^{-(U(q^{\dagger})-U(q_{A}))}
\end{equation}
where $q$ is the reaction coordinate and $q_{A}$ and $q^{\dagger}$ are the locations of the initial well and the barrier respectively. 

In the case of the M\"uller-Brown potential, we assume the ideal reaction coordinate to be along the Minimum-Energy Path obtained using a Nudged Elastic Band method.\cite{henkelman2000climbing,henkelman2000improved,henkelman2001methods} With the potential energy $U(q)$ computed along this path $q$, we use quadratic fits around the initial well ($q_{A}$) and around the largest barrier ($q^{\dagger}$) to find the double-derivative terms. Finally we use this approximate value of $\langle h_{\Gamma}\rangle_{p}$ in Equation (\ref{eq:cgfform}) and (\ref{eq:fprobab}) to obtain the CGF.

\end{document}